\documentclass[preprint,12pt]{aastex}

\newcommand{\kepler}{\textsl{Kepler }}

\newcommand{\etal}{et~al. }
\def\parcmin{{\tt '}\mskip -6.0mu.\,}

\begin{document}

\title{The Burrell-Optical-Kepler-Survey (BOKS) I: Survey Description and 
Initial Results}

\author{
John J.\ Feldmeier\altaffilmark{1}, 
Steve B.\ Howell\altaffilmark{2},
William\ Sherry\altaffilmark{3},
Kaspar\ von Braun\altaffilmark{4},
Mark E.\ Everett\altaffilmark{5},
David R.\ Ciardi\altaffilmark{4},
Paul\ Harding\altaffilmark{6},
J.~Christopher\ Mihos\altaffilmark{6},
Craig S.\ Rudick\altaffilmark{6},
Ting-Hui\ Lee\altaffilmark{7},
Rebecca M.\ Kutsko\altaffilmark{1},
Gerard T.\ van Belle\altaffilmark{8}
}

\altaffiltext{1}{Department of Physics and Astronomy, Youngstown State
University, Youngstown, OH 44555; jjfeldmeier@ysu.edu}
 
\altaffiltext{2}{National Optical Astronomy Observatories, Tucson, AZ 85726} 

\altaffiltext{3}{National Solar Observatory, Tucson, AZ 85726} 

\altaffiltext{4}{NASA Exoplanet Science Institute, California Institute of Technology, Pasadena, CA 91125}

\altaffiltext{5}{Planetary Science Institute, 1700 East Fort Lowell Road, Suite 106, Tucson, AZ 85719, USA}

\altaffiltext{6}{Department of Astronomy, Case Western Reserve
University, 10900 Euclid Ave, Cleveland, OH 44106}

\altaffiltext{7}{Department of Physics and Astronomy, Western Kentucky University, 1906 College Heights
Blvd \#110077, Bowling Green, KY 42101}

\altaffiltext{8}{European Southern Observatory, Karl-Schwarzschile-Str. 2, 85748 Garching Germany}

\begin{abstract}

We present the initial results of a 40 night contiguous ground-based campaign of time
series photometric observations of a 1.39 deg$^{2}$ field located within the NASA 
\kepler mission field of view.  The goal of this pre-launch survey was
to search for transiting extrasolar planets and to provide independent variability
information of stellar sources.  We have gathered a data set containing light curves of 54,687 
stars from which we have created a statistical sub-sample of 13,786 stars between $14<r<18.5$ 
and have statistically examined each light curve to test for variability.  
We present a summary of our preliminary photometric findings including the overall level and content 
of stellar variability in this portion of the \kepler field and give some examples of unusual
variable stars found within.  We present a preliminary catalog of 2,457 candidate variable stars,
of which 776 show signs of periodicity.  We also present three potential exoplanet candidates, 
all of which should be observable in detail by the \kepler mission.  
\end{abstract}


\section{Introduction}

Stellar variability studies provide critical access to a number of
astronomically significant properties including rotation rates,
eclipsing binaries, and pulsations.  Aside from offering insight into
the nature of the stars themselves, statistical studies conducted on
large samples of field and cluster stars broaden our understanding of
stellar evolution and can assist in the search for exoplanets.
Variability surveys that endeavor to find variations with intermediate
periods of days to weeks studying objects
such as short period eclipsing binaries and planetary transits,
benefit from dedicated long-term, high cadence, dedicated observations
\citep[][and references therein]{howell2008,vonbraun2009b}.

NASA's \kepler mission \citep{keplerreview}, which was launched in
March 2009, is conducting a transit search in Cygnus with the goal of 
finding Earth-like planets orbiting in the habitable zones of sun-like stars.  The
signatures of transits of this nature would have relatively small
depths, making the inherent variability of the host star even more
relevant.  The success of this mission partially depends on an accurate
characterization of the stellar variability in the field.  To that
end, the Burrell-Optical-Kepler-Survey (BOKS) was designed to
determine the level and type of stellar variability in a small ($\approx$ 1\%)
subsection of the \kepler field.  As an added goal, we can assess the frequency 
of close-in Jovian-type planets (the so-called ``Hot Jupiters'') in the same field, 
and allow for a comparison of ground-based and the \kepler based transit surveys. 
A number of other Hot Jupiters have been discovered in the \kepler field prior to launch
from ground-based surveys \citep{odonovan2006,pai2008,bakos2010} and \kepler
itself has already discovered many additional exoplanets 
\citep{kepler4,kepler5,kepler6,kepler7,kepler8,keplerbig,steffen2010}.  Further comparisons
of ground-based and space-based transit candidates would be extremely beneficial due to the 
high quality lightcurves that \kepler can provide.  In particular, since the \kepler
mission must make numerous selection cuts in order to achieve their mission objectives
\citep{batalha2010}, there is a tendency to avoid fainter target stars that may have detectable
Hot Jupiters, but not be suitable for Earth-sized transit searches.  By identifying additional
Hot Jupiter candidates, and then having \kepler undertake follow-up observations, we may be able to 
study detailed properties of these systems and characterize them in exquisite detail.  This
has already been done for one pre-launch Hot Jupiter candidate, HAT-P-7, and there are indications 
that the extrasolar planet in this system is gravitationally distorting the host star \citep{welsh2010}. 

This paper introduces the properties of BOKS, and gives a summary
of the data reduction and analysis of the survey.  In \S\ref{sec:obs} we present 
our observing strategy and a summary of our observations.  We outline our data 
reduction techniques and discuss the observational window function of our survey in \S\ref{sec:datared}.  
We present the object detection, photometry and astrometry in \S\ref{sec:objphot}.  Finally, in \S\ref{sec:analyze} 
we discuss the initial results of our variability survey and of the search for exoplanets in the 
BOKS field, specifically those of the ``Hot Jupiter'' variety.  We review our conclusions 
in \S\ref{sec:conclusions}.  There are a number of other scientific 
projects planned for the BOKS data, such as the comparison of stellar field to cluster variability, 
cataloging the many variable stars found in this survey, and searching for any moving objects.  
These results will be discussed in future papers.  We plan to submit all of the BOKS 
data to the NASA/IPAC/NExScI Star and Exoplanet Database\footnote{located at: http://nsted.ipac.caltech.edu/} 
\citep{vonbraun2009a}, where it can be of service to the entire astronomical community

\section{ \label{sec:obs} Observations}

In order to maximize the scientific benefits from this survey, we chose our scientific
field under a number of constraints.  First, besides being located within the \kepler field 
our field ideally should have a large number of stars, as this will improve our chances
to find extrasolar planets and find other interesting objects.  Second, given that our 
ground-based imager has relatively large pixels in angular size, the field must not be so 
close to the Galactic plane that photometric crowding would be a major factor.  Third, in 
order to compare our data against numerous stellar cluster variability surveys, such as 
UStAPS\citep{hood2005}, EXPLORE/OC \citep{vonbraun2005}, PISCES \citep{moch2006}, and STEPSS
\citep{burke2006}, we decided on a field that had both field stars and an open cluster within it, so 
that we could compare the variability properties of both stellar populations simultaneously.  
In order to find the optimal field, we first pre-imaged a number of candidate fields on 
2006 April 24.  After visual inspection of all of the pre-images, we made a determination
of the field that best matched our conflicting criteria.

Our final selected target field of view covered 1.39 deg$^{2}$ in the constellation of 
Cygnus and was centered on the open cluster NGC 6811 (RA$=19^{h}37^{m}17^{s}$,
Dec=$+46^{d}23^{m}18^{s}$; WEBDA\footnote{The WEBDA database, developed by J.-C. Mermilliod, 
can be found at http://www.univie.ac.at/webda/}).  Our field is completely contained 
within the \kepler field, and the BOKS field is located on channels 63, 47, 23, and 39
of the \kepler imager in the Spring, Summer, Fall, and Winter seasons respectively
\footnote{Full Frame Images (FFIs) of the Kepler field for each observing season can be found at
 http://archive.stsci.edu/kepler/ffi\_display.php}.
Our photometric survey began on 1 September 2006 and ended on 10 October 2006, 
consisting of 40 nights in all.  We observed with the Case Western Reserve University 0.61m Burrell 
Schmidt telescope (hereafter the Burrell), located at Kitt Peak National Observatory. One advantage of a 
dedicated observatory for this survey is that we could observe for a large number of consecutive
nights.  Many researchers
\citep{pont2006,beatty2008,vonbraun2009b} have shown that the total duration of a photometric survey 
is crucial in maximizing transit detection efficiency in the presence of statistically correlated (``red'') 
noise.

The imager used for this survey was a SITe back-illuminated 2k$\times$4k CCD 
with 15 micron (1.45 arcsecond) pixels, run by a Leach version 2 controller \citep{leach1998}, and
two output amplifiers.  The long axis of the CCD was oriented East/West.  We observed primarily in 
the SDSS $r$-band filter, but we also obtained occasional Johnson $V$-band images of the field of 
view in order to obtain two-color information and to allow for
cross-comparison between our data and other photometric catalogs.
In particular, we compared our photometry to that found
from the Kepler Input Catalog (KIC)\footnote{Version 10 of the KIC is available at: 
http://archive.stsci.edu/kepler/kic10/search.php In some cases, we used the 7th and 8th version of the KIC for 
steps in our analysis.  These will be referred to as KIC78 in the text.}

\begin{figure}
\figurenum{1}
\plotone{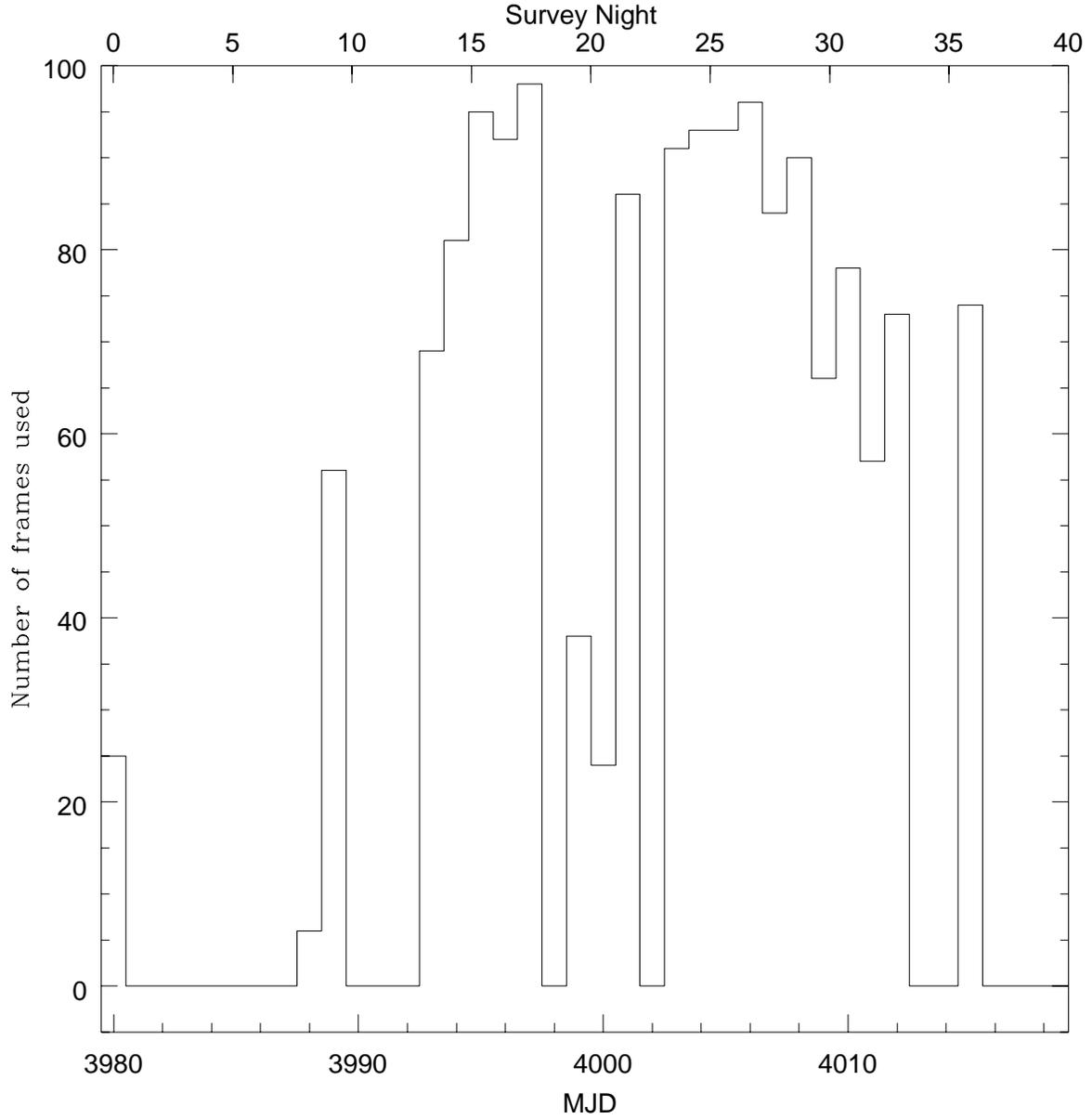}
\caption[ Number of Images ]{\label{fig:hist} The number of BOKS images
taken as a function of survey time.  Due to the effects of thick clouds, and occasional
high winds, the numbers of nightly frames taken vary significantly throughout the survey.}
\end{figure}

Observations were ongoing during any weather conditions where stars were
visible on the sky, and it was safe to operate the telescope.  As a result, the BOKS data
have large variations in seeing, transparency, and night sky brightness.  Of the 40 nights of
observing, thirteen were completely lost to weather, leaving 27 nights of potential
data.  Nearly all of our $r$-band and $V$ integrations were 180 seconds in duration, with the 
exception of 31 images taken on night 14 that had exposure times of 300 seconds.  The
CCD readout time was 45 seconds in length.  A total of 1,924 $r$ images and 10 $V$ images were taken 
over the entire run.  The SITe CCD gain was fixed at 2 electrons per ADU, each pixel had
a full well capacity of at least 100,000 electrons, and the read noise was 12 electrons.
We note here that while the CCD pixel scale is fairly large compared to typical CCD
imagers, studies (Feldmeier \etal 2011, in preparation) have shown that
milli-magnitude relative photometry is possible, even with such large pixels.  Table~\ref{table:obslog}
provides a summary log of our observations and Figure~\ref{fig:hist} shows a graphical representation of
the number of exposures throughout the survey.  An $r$-band exposure of BOKS, created from 
co-adding the first 24 images from our survey is shown in Figure~\ref{fig:field}.

With the help of the American Association of Variable Star Observers (AAVSO), we also arranged to have 
bright variable stars photometrically monitored in the field at the same time as BOKS was underway.  The 
preliminary results from this independent photometric survey are discussed in \citet{henden2006}, and a more 
careful comparison will be discussed in a future paper\footnote{The AAVSO NGC~6811 campaign information 
can be found at\\http://www.aavso.org/news/ngc6811.shtml}.

\section{ \label{sec:datared} Data Reduction }

Although the data reduction of BOKS is relatively straightforward, the large
number of images and the need to ensure highly precise relative photometry demands 
some careful attention.  We therefore began our 
CCD reductions as follows.  Since our imaging observations were obtained using the dual 
amplifier mode with the SITe CCD,  we first combined the two amplifier readouts into single 
images using IRAF's\footnote{IRAF is distributed by the National Optical Astronomy 
Observatory, which is operated by the Association of Universities for Research 
in Astronomy, Inc, under cooperative agreement with the National Science Foundation.}
\textsf{mscred.mkmsc} task.  The resulting images were then merged, trimmed, and overscan 
subtracted using the \textsf{ccdred.ccdproc} task.  

We obtained approximately ten bias and five twilight flatfield frames per night, which
were used on the respective night's images using IRAF's
\textsf{ccdred} package after they were checked for unwanted features
such as bright stars in the flatfields or amplifier noise in the bias
frames. If any such features were present, the corresponding bias or
flatfield frames were discarded.  We created nightly master bias frames, but
due to the presence of dust grains on the dewar window and/or filter, that changed 
positions between nights, we did not create a master flat for the entire run.  

After the data had been processed, we inspected the entire data set for quality
issues.  We determined basic parameters of each image such as the median sky level
and the median seeing in order to verify that our data was suitable for use.
To determine the median sky level, we used the IRAF task \textsf{imstatistics} in iterative mode.
The median seeing of each image was calculated by applying the \textsf{imexam} task on
150 bright, unsaturated stars on each image.  Each star was fit using a Gaussian function, and 
the median of all of the derived FWHM values was adopted for the median seeing for each image.

After applying the photometric zero point (\S 4.3), we present the median sky brightness
and median seeing for BOKS in Figure~\ref{fig:skysee}.  As would be expected from a 
telescope run of this length, these properties varied substantially as differing lunar phases 
and weather patterns occurred.  After this analysis, we removed night 37's data due to very 
high sky levels by the nearly full moon.  Next, we visually inspected each image for 
quality issues.  From this
process, we found that night 12 suffered from condensation on the CCD dewar window, leaving a 
circular distortion feature on approximately 17\% of the area of each image.  Night 19 suffered from 
CCD electronics issues on one of the two CCD amplifiers.  In both of these cases, many of the stars 
remain unaffected on each image.  However, to be conservative in this initial study, we have 
removed the entire night's data from further consideration.  After our initial light curve
analysis (discussed in \S\ref{sec:analyze}), we found that two additional nights, nights 13 and 23, had extremely
variable clouds (with such a large angular field, even the ensemble photometry method discussed
in \S\ref{sec:objphot} can fail if the clouds are variable enough; Feldmeier \etal 2011, in preparation).  
Although some of the exposures on these nights should be acceptable for our scientific goals, we again chose 
to be conservative and removed the entire night's data from consideration.  This left us with imaging 
data from 22 different nights over the span of the survey and a total of 1,565 $r$-band images that could be 
used for potential variability studies.  

For a planetary transit survey, understanding the observational window function gives crucial insight
into the survey completeness and sensitivity to periodic variability.  Normally, the window
function is defined as the probability that a planetary transit is detected in a given data set, as 
a function of planetary orbital period \citep{vonbraun2009b}.  For BOKS, we calculated the 
approximate observable window function in the following manner: we simulated planetary transits 
from periods ranging from 0.5 to 30 days 
and divided each period into 10,000 phases.  Given the starting and ending times of the 22 nights of 
observations, we then determined the probability of observing a transit with that period over all phases.  
The results of this analysis are plotted in  Figure~\ref{fig:window}.  As with all time-limited photometric surveys, 
we are most sensitive to short period transits, and our ability to detect transits decreases with transit period.  
Since our survey is ground-based, the characteristic aliasing period of integer days is also strongly present 
in our data.  We should note that this window function deals with temporal sampling only 
and likely to be an overestimate: it does not take into account the effects of differing transparency, 
seeing, and sky brightness on the detectability of transits.  It also does not take into account the effects 
of differing stellar and planetary radii on the detectability of transits.  Finally, the effects of
statistically correlated ``red noise,''  which are likely to be significant, are not included in this 
calculation.  We plan to perform extensive Monte Carlo simulations on these effects, which will be presented 
in a future paper.  However, we note that BOKS has a significant advantage over many other ground-based 
transit surveys: if even one transit appears in our survey, we could, in principle, verify it with \kepler 
follow-up observations.

\section{\label{sec:objphot}Object Detection, Photometry, and Astrometry}

With our good quality data set finalized, we next focussed our attention on finding all stellar 
sources in the BOKS field, and determining their magnitudes and positions throughout the survey period.  

\subsection{Object Detection and Coordinate Transformations}
We next created a master image by combining six 
individual images from night 25 using the \textsf{imcombine}
task.  This master image was used to ensure that we detect all of the stellar sources in the frame and 
remove the possibility
of radiation events contaminating our source catalog.  We next created the master list of stars (point sources) 
by running the \textsf{daofind} task within IRAF on the master image.  We chose a \textsf{threshold} value 
for point source detection of five times 
the standard deviation of the sky background.  Given the large pixel scale of our data, we adjusted
the \textsf{sharphi} value to 0.9 rather than the 1.2 that is normally adopted.  All other data-independent 
parameters were left at their default values.  We found a total of 56,354 sources that matched the \textsf{daofind} 
criteria we adopted.  This initial list was then manually inspected to ensure that non-stellar objects were not 
included, such as diffraction spikes, radiation events, or objects on the extreme edges of the frame.  This left 
a total of 54,687 objects for further study. 

With the master list of coordinates determined, we then needed to re-identify each source on every frame
of our survey.  Rather than shift each image, which would lead to unacceptable uncertainties in the magnitudes due
to interpolation, we instead re-determined the source coordinates for each individual image.  To do this, we used a high
quality image from the middle of night 25 as our positional reference image. To match stars in individual images to 
those found on the reference image, we split each image into eight rectangular subsections (512$\times$512 pixels).
For each subsection of each image, we summed a group of rows and group of columns to create a pair of one dimensional 
arrays that contained the peak counts at the row or column location of each star.  We then used the Fourier transform
between pairs of these arrays taken from different images to locate translational shifts between each subsection of 
the images.  These shifts were applied to the master coordinate list to locate each star on individual images.  
The Burrell tracked fairly well overall but in periods of cloudy conditions, when the autoguider was
unable to hold the tracking steady, the spatial shifts could be up to 1--2 arcminutes.  By dividing the image into 
subsections, we were able to accommodate most of the magnification and rotational differences between images 
in our final astrometric coordinate solution.

\subsection{Aperture Photometry and Ensemble Correction}
We performed aperture photometry using the \textsf{phot}
task within IRAF's \textsf{noao.digiphot.daophot} package \citep{stetson1987}. We experimented with
five seperate photometry apertures, from very small radii (1 pixel) to large radii (5 pixels), to 
span the range of stellar brightness and seeing changes over the observed field of view.  After inspection 
of the output light curves, we selected two distinct aperture values for our 
work (3 and 4 pixels) based on their small level of scatter around the median magnitude of a representative 
light curve sample. These two aperture values, which correspond to 4.35 arcseconds and 5.8 arcseconds, with a 
sky annulus of 11.6--33 arcseconds, gave good results and we used the smaller aperture for our final light curve set.

Once the raw photometry files were created, we then needed to account for the effects of seeing, 
transparency, and airmassfor each star.  To do this, we adopted a local ensemble 
photometry technique, where we used a local set of bright stars that 
are photometrically constant to determine changes in these parameters.  The algorithms are discussed in detail 
by \citet{everett2002} and \citet{everett2001}; we will briefly outline them here. 

First, we divided each image up into 8 $\times$ 4 square regions with a size of 500 $\times$ 500 pixels (corresponding
to 725 arcseconds on the side). This size was chosen to allow a sufficient number of ensemble stars to be present in
the individual regions and at the same time, to optimize our sampling of positional dependence of photometric 
effects such as variable point spread functions, color terms, or focus gradients.  For a star to be an 
acceptable ensemble star, it must fulfill certain quality criteria. Specifically: 1) the star must be present in 
all frames, 2) it must have a photometrically constant light curve ($\chi^{2} < 3.$), 3) it must be bright enough that 
photon noise is negligible (average flux must be greater than 50,000 ADU, corresponding to a S/N of 316 or better), 
and 4) have no close-by stellar companions that would interfere with the light curve in poor seeing conditions (no stars
within 5 pixels that are within 5 magnitudes of the ensemble candidate's magnitude).  We 
created an initial list of ensemble stars by conducting an automated search through the stars in each region.  
After this automated preselection, the ensemble stars were inspected by eye to eliminate stars that appeared to have 
signs of residual variability compared to the remainder of the ensemble stars.  A total of 688 stars with $r$ magnitudes
varying between 14.3 and 16.6 were used in the final ensembles. 
 
After the final list of ensemble stars was created, the relative photometry procedure was rerun using 
only the cleaned sample of stars.  If a region had fewer than ten ensemble stars, we combined it with 
a neighboring one to create a larger region.
The exact calculation of relative photometric offsets for the individual regions 
was performed by a custom written routine based on \cite{everett2002}.  An example of these offsets for a single
region is plotted in Figure~\ref{fig:offsets}.  Due to the effects of airmass, seeing, and transparency, these
offset values vary significantly over the survey length.

Due to the differing positions of each individual image, various stars have differing numbers of observations, with 
objects at the edges of the fields having fewer observations than object near the center.  To ensure a high quality 
set of light curves for study, we focus exclusively on light curves that have at least 1,000 photometric measurements.
This left 32,806 sources for further analysis.

An unfortunate issue we found during the data reduction was the fact that the start time (HJD) for 
any given exposure listed in the image header was not sufficiently accurate for our purposes.  
The internal clock on the data recording computer, which was a Microsoft Windows PC running the Voodoo
image acquisition software\footnote{located at http://www.astro-cam.com/} at the telescope was found to be 
imprecise and could not be used alone for exact timing purposes as we found that it drifted by up to 
several minutes over the course of a single night. As a result, we used the file creation times 
(recorded by a different clock on an internet time controlled Linux machine) as HJD ``start time" 
information in our image headers.  From some experimentation, we found that the header start time recorded is 
approximately 93 seconds after the true mid-exposure time in most cases.  From comparing consecutive exposures
throughout the survey, the uncertainty in this correction is approximately one second.
Consequently, the times we recorded in our light curves will be precise relative 
to each other, but our time zero point is not tied to UT or any other absolute time system to within several
minutes.  In the future, we plan to correct this by correlating our observations against the AAVSO 
and \kepler observations, which should allow us to reduce any time offsets.

\subsection{Photometric Zero Point}

The transformation from differential instrumental magnitudes to SDSS $r$ magnitudes for our stars was 
performed by comparing the 
magnitudes of the ensemble stars in each subregion to the corresponding $r$ stellar magnitudes listed 
in the Kepler Input Catalog (KIC).  The scatter in this comparison was approximately 0.1 magnitudes, which we 
take as our systematic magnitude uncertainty for the BOKS survey.  Given that we have minimal color
information in our observations and our survey is primarily interested in searching for stellar variability, 
this amount of uncertainty is acceptable for our needs.  For any individual object, however, 
one can derive a more accurate magnitude by directly comparing the object to the KIC values, which have a 
mean $r$ systematic uncertainty of less than 0.015 magnitudes\footnote{photometric uncertainties for the KIC 
can be found at: http://www.cfa.harvard.edu/kepler/kic/kicindex.html}.

\subsection{\label{sec:ast}Astrometry}

We performed astrometry using the USNO-B1.0 catalog \citep{monet2003} and 
\textsf{wcstools}\footnote{\url{http://tdc-www.harvard.edu/software/wcstools/}}.  In order to eliminate the 
effects of field rotation and distortion, we performed astrometry on the individual regions as described 
above, rather than on the field as a whole.  Our internal uncertainties on our coordinates were approximately
$\pm$ 0.5\arcsec right ascension and $\pm$ 0.3\arcsec declination.  As an independent check on our astrometry, 
we compared our final astrometric catalog to that found from the 2MASS Point Source Catalog \citep{2mass}.  
Figure \ref{fig:match} shows the astrometric comparison between our BOKS astrometric solutions and those
listed for common stars in the 2MASS catalog. The one-sigma coordinate offset is near 0.4\arcsec both 
RA and DEC with a small, but clear, asymmetry across the field.  Given that the pixel size of the 
Burrell is 1.45\arcsec and that the wide field of the Burrell produces differential focus and refraction
effects, this agreement is quite reasonable.

\section{ \label{sec:analyze} Light Curve Analysis }

With the light curves established for each star, we then began the search for photometric variability.  
To characterize the possible variable nature of our light curve sample and to search for possible 
exoplanet transits, we utilized OPTICSTAT, a custom-written statistical 
analysis package created by M. Everett (discussed in Howell \etal 2005).  OPTICSTAT 
returns several statistics related to stellar variability, including the reduced ${\chi}^2$ 
(${\chi}^2_{\nu}$), the standard deviation, the probability that the star varies periodically, and the
most likely period.  Prior to our statistical light curve analysis, we removed the
effects of obvious cosmic rays which will artificially increase the apparent flux as follows.  
If there were one or two
consecutive points in the light curve that deviated from the mean
magnitude by more than 3.5 times the standard deviation, those 
points were rejected during statistical tests.  This might remove some signs of 
ultra-short variability, but since the exposures are 180 seconds in length, and the
readout time of the CCD is 45 seconds, the total scientific impact should be minimal.

To test each light curve for general variability, we fitted the
light curve with its mean flux and then calculated the probability that the
reduced $\chi_{\nu}^{2}$ statistic shows the data to be inconsistent with
this mean flux.  This test can easily fail, however, in the presence
of systematic errors or uncertainties in the calculated magnitude
errors we assign to the individual data points.  To compensate, we adopted a more
conservative threshold for the $\chi_{\nu}^{2}$ probability than our
formal errors would dictate.  When applied to the full light curve, 
we have adopted the criteria that point sources that 
have $\chi_{\nu}^{2}\geq5$ are variable sources.  

\subsection{ \label{sec:errors} Photometric Uncertainties}

In order to make a proper determination of variability for each star, we must determine
the random and systematic uncertainties for our target stars.  By splitting the entire 
BOKS field of view into 32 smaller sub-sections we
were able to remove the vast majority of systematic variations in our light
curves.  However, some dispersion remains due to photometric uncertainties,
small differences in color and properties between the ensemble stars and the
target objects, and the statistically correlated (``red'' noise) that is 
present in all ground-based transit surveys \citep{pont2006,beatty2008,vonbraun2009b}.

To estimate our remaining photometric uncertainties for every light curve, we 
calculated a weighted average magnitude and the standard deviation of the entire light curve
about this average.  Figure \ref{fig:sigma} shows the standard deviation of our BOKS light 
curves as a function of their $r$ magnitude.  As can be clearly seen, our brightest
stars have a 4 milli-magnitude dispersion about the average magnitude, which sets our noise floor
for this survey.  For stars brighter than $r\simeq14$, saturation becomes an issue with seeing 
changes making the exact magnitude of saturation somewhat imprecise.   

As the apparent magnitude increases, the photometric errors also increase.  Plotted on Figure~\ref{fig:sigma}
is an estimate of the photometric errors as a function of target flux \citep{everett2001}:
\begin{equation}
\sigma_{*} = 1.0857 \frac{\sqrt{N_{*} \times g + n_{pix} [ 1 + (n_{pix}/n_{sky})](N_{sky} \times g + R^{2})}}{N_{*} \times g}
\end{equation}
where $N_{*}$ is the number of ADUs from the star in the aperture, $g$ is the gain of the CCD in electrons
(2 e$^{-}$ per ADU), $n_{pix}$ is the number of pixels in the aperture, $n_{sky}$ is the number of 
pixels in the annulus around
the aperture used to measure the sky flux, $N_{sky}$ is the flux in ADUs per pixel from the sky, and $R$ is
the rms read noise of the CCD in electrons (12 e$^{-1}$).  For this comparison, we assumed the faintest 
value of the night sky brightness in our survey, which should give a lower bound to the true photometric 
uncertainties.  As can be seen, this function is in good agreement with the lower edge of our 
error distribution.  At the faint 
end of our photometry, near $r=19$, our uncertainties are $\sim$50 milli-magnitudes per observation, 
substantially larger than the 10-30 milli-magnitude precisions needed to find transiting 
extrasolar planets.

For completeness, we have also calculated the variability level expected from atmospheric 
scintillation alone (Young 1967; see also Young 1993a; Young 1993b; Badiali et al. 1994 
for some important comments) in order to rule it out as a significant contributor of our highest
precision photometric results.  We find that the scintillation
at unit airmass is approximately 0.3 milli-magnitudes, about
a factor of 15 lower than the photon noise from the brightest
stars in our sample.  This is an approximate value: Young (1993b) notes
that the value can vary by up to a factor of two on a timescale of a
few minutes.  Nevertheless, given our measured photometric uncertainties, scintillation is not a major
contributor to our error.

\subsection{Point Source Variability Statistics}
\label{sec:vari}

An important goal of BOKS is to determine the variability properties of stellar sources in general
within one portion of the \kepler field.
As mentioned previously, DAOFIND identified 56,354 point sources in at least one BOKS survey image.  
Of these, 32,806 point sources were observed at least 1,000 times within our survey.  From this subset, 
25,861 point sources are located within 2.5 arcseconds of a source in the KIC78, and therefore we have 
additional photometric information.  However, since the KIC78 data is non-contemporary with BOKS, any
values derived from this comparison should be treated carefully.  For statistical purposes, we
excluded any source from the KIC78 that did not have a valid $r$ magnitude, even if it had measured 
magnitudes in other bands.  This left 22,340 sources for further analysis.

We then determined a magnitude cut-off value for our statistical analysis of variability.  
Progressively fainter stars have larger photometric errors, are more likely to have poor quality light curves 
due to contamination from nearby bright stars, and in conditions of poor seeing, and may have incorrect recentering 
by the aperture photometry algorithm, which can cause the aperture to recenter on a nearby brighter star.  
For these reasons, we restricted our variability statistics to the 13,786 stars brighter than $r=18.5$, 
which also lie within 2.5\arcsec of a source within the KIC78 and which have over 1000 data points.  We 
hereafter refer to this subsample as the BOKS-KIC sample.

After applying the statistical tests from OPTICSTAT,  we found 2,457 stars 
with $\chi_{\nu}^{2}\geq5$ and $r<18.5$ in the BOKS-KIC sample.  We note that this number is approximate,
as the reduced $\chi^{2}$ is strongly affected by a number of systematic uncertainties, such as residual
variability in our ensemble stars, color mismatches between the ensemble and the target stars, spatial
structure in the extinction correction, and other effects.  Additionally, photometric uncertainties 
and stellar variability applied to the KIC78 will move stars above and below the magnitude cutoff, creating 
an additional systematic uncertainty.

The variability fraction found ($\approx$ 18\%) can be compared to other variability surveys
using identical search techniques.  In an earlier study, \citet{everett2002} found a variability fraction
of 17\% over a five day survey period using a similar telescope and sampling rate, but using a less
strict variability criteria ($\chi_{\nu}^{2}\geq3$).  In contrast, the variability study of NGC~2301
\citep{tonry2005,howell2005,sukhbold2009} found a much larger variability fraction (56\%) using an orthogonal
transfer CCD observing mode over a 12 night run, and with substantially better photometric precision
($\approx$ 1.6~mmag).  
  
\citet{tonry2005} and \citet{howell2008} discussed a relation between the percentage of variable
sources that will be found in any given photometric survey and that
survey's photometric uncertainty.  From the NGC~2301 results, the cumulative fraction of variable
stars found, as a function of quartile variability, $x$, (the quartile variability is $\approx$ 1.5 times
smaller than the standard deviation) is:\footnote{Again, we note that in this paper we use $\chi^{2} > 5$ while the 
previous studies used $\chi^{2} >3$, which
would allow more low amplitude variables into the samples.}
\begin{equation}
f(<x) = 1 - \frac{1.6~mmag}{x}
\end{equation}
For a best precision of 4 mmag, as we have in the BOKS survey, the percentage of variable sources would
be expected to be $\sim$40\%, substantially larger than what was actually detected.  Some of this difference 
may be due to differing stellar populations:  NGC~2301 is a young open cluster 
\citep[250 Myr;][]{kim2001}, in contrast to NGC~6811 \citep[575 Myr;][]{luo2009}, and the level of stellar 
activity on the main sequence may be significantly different between the two clusters.  The amount of
background and foreground contamination may vary significantly between the two surveys as well.  Given
the position of the BOKS field, it is likely that many of the stars are field stars and therefore have
lower amounts of variability.

Of the 2,457 variable sources, 776 (32\% of all the variables or $\approx$ 6\% of the total BOKS-KIC sample) 
were found to be periodic variable candidates (see \S\ref{sec:pervar}).  This is significantly larger than 
the results of \citet{everett2002}, who found a 14\% periodic fraction in their survey and \citet{howell2005}, who found
a 13\% periodic variable fraction in NGC~2301.  We believe this is due to the substantially longer 
time coverage of BOKS, which should make it substantially easier for our search algorithm \S\ref{sec:pervar} 
to detect periodicity.  The remaining variable stars appear to be non-periodic within the limits of 
our time sampling, photometric precision, and observational window.  These non-periodic variable 
stars are presented in Table~2, with the corresponding photometric information from the KIC given.  We
caution that the standard deviations given for these light curves are likely to be an overestimate: 
any photometric residuals in the ensemble stars and the effect of radiation events can increase this
value significantly. 

There have been three other recent variability surveys of the regions of the BOKS field, all centered
on the NGC~6811 open cluster \citep{vanc2005,rose2007,luo2009}.  Unfortunately, the majority of the stars 
found to be variable by these surveys are saturated on the BOKS images.  The three exceptions, 
stars V8 and V9 found by \citet{vanc2005}, and star V17 found by \citet{luo2009}, were also 
detected by BOKS, and all three were found to be periodic variable stars.

\subsubsection{Non-variable Stars}

Our variability analysis found 10,881 stars brighter than
$r=18.0$ with no detected variability (${\chi}^2_{\nu}<$5).
Figure~\ref{fig:const_grr_cmd} shows a color-magnitude diagram
(CMD) of these stars.  This CMD is typical of stellar fields
within the galactic disk, showing a plume around a color
of $g-r = 0.4-0.6$, corresponding to the main sequence turn-off
for an old ($>$ 10 Gyr) stellar population.  The BOKS survey combined with 
the KIC does not go photometrically deep enough to detect the very faint and red low-mass
stars in the disk population, which typically appear at
$g-r  \approx 1.4$ and begin at $r \approx 18.2$.  Similar CMDs are found in 
the work of \citet[][see their Figures 2 \& 3]{dejong2010} with data taken 
from the Sloan Extension for Galactic Understanding and Exploration (SEGUE).  

Figure~\ref{fig:const_n_vs_r} shows a histogram of the number of
photometrically constant stars as a function of $r$ magnitude.  The number of photometrically
constant stars increases roughly linearly up to the final bin ($r=17.9$).  This
reflects both the increase in the number of faint stars and our
decreased sensitivity to variability for fainter stars. 

Some example light curves of stars that show little signs of variability over the 
timescale of our survey are given in  Figure~\ref{fig:constar}.  Note that in such long 
surveys, it is highly likely for a star to suffer at least one, and possibly 
more, hits from radiation events, such as cosmic rays.    

\subsubsection{Periodic Variable Stars}
\label{sec:pervar}

In order to test each variable star light curve for periodicity we applied the Lomb-Scargle
method \citep{lomb1976,scargle1982}, as described by \citet{press1992}.  
The algorithm produces a periodogram giving probabilities for sinusoidal signals 
in the light curves over a range of periods from our minimum sampled period of
$1/60$ days ($0.4$ hours), set by the spacing of consecutive exposures, up to 40 days, 
the maximum possible period spanned by the entire data set.  We identified stars 
as possible periodic variables if: 1) OPTICSTAT returned a false probability of 
periodic variability of less than  $1\times 10^{-4}$, 2) the amplitude was less 
than 2.5 magnitudes, and 3) the average magnitude was brighter than $r$=18.5.  
This yielded a large but manageable list of candidates.  We then visually inspected the light
curves of these candidates after phasing them to the best-fit period.  
Only those light curves that had clear periodic signals, and whose light curves
showed no sign of systematic effects were accepted as potential periodic variable stars.
In order to reduce the effects of aliasing, we also removed from our sample any sources that
had periods of 1 $\pm$ 0.025 days, though some of these objects may be genuine variable stars.

We found that 776 stars from our variable sub-sample had periodic signals that
ranged from $\sim$0.2 day to $\sim$33 days.  The coordinates, mean $r$ magnitudes,
determined period and amplitude and the corresponding KIC information are presented in Table~3.
Additionally, we have classified the periodic variables into approximate types, which are
also presented in Table~3.  Of these objects, 78 ($\approx$ 10\%)
show variability amplitudes larger than 
$\sim$0.1 magnitudes.  Another 235 objects ($\approx$ 30\%) have periodic amplitudes of 1-3\% 
and periods of 1-3 weeks, which is consistent with rotational modulation due 
to star spots.  A significant number ($N=93$; 12\%) of the periodic variables
remaining have periods less than 2~days and photometric amplitudes less 
than 0$^m$.05.  These short period low amplitude variables are likely to be 
pulsational variables such as $\delta$~Scuti stars \citep{breger2000}.

We next compared the properties of the periodic variable stars as a function 
of period, stellar color, and magnitude, as these distributions give insight 
on stellar properties in general.  It is well known that there are systematic
changes in the fraction and amount of stellar variability across the H-R
diagram \citep[][and references therein]{eyer2008,ciardi2011}, but the precise distributions
are still under debate.  Second, studies of these distributions allow
us to compare our results with other high precision variability surveys and provide
confidence that the survey is valid.  Figure~\ref{fig:logp} shows the overall distribution of 
periods in our sample of periodic variables.  The most notable features of this plot are the
large number of stars with periods longer than 10 days and the peak
in the period distribution near periods of $\sim$1~day.  Despite our attempts to 
remove them, this peak strongly suggests that a fraction of these variable stars have derived 
periods that are reflective of aliasing due to our observational window function, 
rather than their actual period.  Additionally,
the stars found to be periodic with the longest periods only have one or two measured
periods, and the period-finding algorithm may have mistakenly flagged these objects, when in
fact, they may not be periodic.  More observations of these particular stars will be required to 
fully address this issue.

In contrast to the number of photometrically constant stars, a histogram of the number
of periodic variables vs $r$ magnitude, which is plotted in Figure~\ref{fig:var_n_vs_r}, 
shows a steep rise in the number of periodic variables between $r\sim14$ and
$r\sim16$, which is followed by a steep decline in the number of periodic stars 
fainter than $r\sim17$.  This decline is due to the rapid loss of photometric 
sensitivity to low amplitude variations for the fainter stars.
Figure~\ref{fig:var_n_vs_gr} shows a histogram of the number of
periodic variables versus $g-r$ color.  The overall shape of this
distribution is very similar to the color-magnitude distribution for constant stars
(see Figure~\ref{fig:const_grr_cmd}).  There may be a small excess of
the bluest stars ($g-r<0.3$) and the reddest stars ($g-r>1.0$), but it is unclear
whether this is statistically significant.
In contrast, Figure~\ref{fig:var_p_vs_gr} plots the period of periodic stars versus
their $g-r$ color.  Most variables have colors that are similar to
those of constant stars regardless of period.  There seems to be a
small excess of stars with $0\leq g-r \leq0.5$ among the periodic
variables with the shortest periods (P$\lesssim$1 day).

Figure~\ref{fig:amplitude} presents the distribution of amplitude for the periodic
variables within the BOKS survey.  As has been previously seen \citep{howell2005}, 
the number of variable stars increases as the amplitude of variability decreases.  
The dashed line indicates the power law model of variability found by \citet{tonry2005}, 
which has a slope of -2.  The fit is in good agreement, giving further evidence to 
applicability of this model.  

Although a full accounting of the BOKS periodic variable stars would be soporific, we 
briefly discuss some of the more interesting stars here, and we discuss two extreme 
examples in \S~\ref{sec:othervar}.  Among the periodic stars with derived amplitudes larger 
than 10\%, we find at least ten clear pulsational variables.  An example light curve 
of this type of variable star is plotted in Figure~\ref{fig:pulsvar}.  We have also detected two 
probable RR~Lyrae stars within the field, with approximate periods of 0.53 and 0.56 days, 
respectively.  A light curve of one of these objects is plotted in Figure \ref{fig:rrlyrae}.

There are at least 32 eclipsing/contact binaries within the BOKS survey field, 
with periods varying from 0.13 to 6.10 days. Figures \ref{fig:eb1} and \ref{fig:eb2}
give examples of two of these systems.  From visual inspection of the light curves, 
the majority of these systems are of the W Ursae Majoris subtype, as would be expected 
\citep{hoffman2009}.  However, we have also found at least eight Algol-type binary systems.  

\subsection{Exoplanet Transit Candidates}
\label{sec:occult}

A primary goal of BOKS is to search for any signs of transiting extrasolar planets in
the data set. To search the light curves for transits, we used a simple test to find
and flag light curves containing at least one occurrence of a
diminution in the relative flux with parameters specified below.  The
algorithm searched a light curve for time intervals when the mean
magnitude is statistically significantly fainter than in the preceding
and following data points, in other words, an inverse ``top-hat'' light
curve.  The algorithm stepped
through time in each light curve testing a grid of possible interval
starting and ending times and durations and reported back the most
significant possible transit found for each
light curve.  A probability of significance is determined for all
light curves for each possible transit by calculating a Student's
t-test \citep{press1992} statistic comparing the mean magnitude during
transit to the mean magnitude of combined pre- and post-transit data
points.  Those light curves with the most significant probabilities,
specifically those that have a formal false probability less than 
$1 \times 10^{-16}$, were then subjected to further inspection.  In order 
to avoid detecting too many spurious light curve fluctuations as transits, 
we next placed further restrictions on the events reported by the transit-finding 
algorithm.  First, we searched for transit durations only between one and 12 hours.  
Second, we placed limits on the depth of transit that merited further study.  
Large planets orbiting all but the smallest M-dwarf stars will result in transit 
depths of 0.5 magnitudes or less.  We therefore removed all possible transits
that had depths larger than 0.5 magnitudes.  

For the purpose of completeness, we decided to use all of the stars in this
analysis.  It is extremely unlikely that faint stars will show a genuine
transit event, but including these objects in the search allows us to test for
other systematic effects in the algorithms.  The entire BOKS sample 
contains 54,687 light curves that were all analyzed using OPTICSTAT.  
Using our detection limit described above, OPTICSTAT identified approximately 
1,445 light curves with evidence of transit events.  

Each of these 1,445 light curves were then inspected by eye with careful attention 
given to additional criteria.  We required at least two data points during 
transit and at least two data points in both the pre- and post-transit light 
curve, and we also required that the pre- and post-transit data points are 
separated by no more than 24 hours from the time of ingress or egress.
The transit search algorithm used by OPTICSTAT searched for an
inverted ``top hat" shape in the light curve.  However, there are a
number of types of variability that can lead to a ``top hat" shape
that must be eliminated through a visual inspection of the light
curve.  Extrasolar planets will cause transits that have a
flat-bottomed appearance, so any sharp-bottomed transits were rejected from further 
consideration.  It is not possible with our data set to observe
secondary transits resulting from the planet passing behind its star,
so any light curve showing secondary transits was also removed from further study.  
Finally, in order to do follow-up observations we required that the light curve
have at least two transit events.  We require this to confirm the
first transit-event and to obtain an accurate determination of the
planet period before any follow-up observations are planned.

The light curve shown in Figure \ref{fig:eb2} was incorrectly identified by OPTICSTAT
as a transit candidate, but was easily removed by our manual inspection criteria: the transits in this light curve 
are too deep ($\approx$ 0.6 magnitudes), sharp-bottomed, and there is an obvious secondary
transit.  Nearly all of the transit candidates detected by OPTICSTAT were rejected using the 
simple requirements we have outlined.

At the end of our analysis we were left with three exoplanet candidates: 
BOKS-45069, BOKS-40959, and BOKS-52481.  Some basic properties of these candidates 
are given in Table~\ref{table:bokscand}, and a plot of their lightcurves is given
in Figure~\ref{fig:candidates}.  For each of the candidates, we then determined an approximate
period by phasing the light curve to the best-fitting phase.  In the case of BOKS-52481, 
we detected one full transit and only a portion of another transit, so the measured period
is substantially less certain than the other two candidates.  We also determined an approximate
transit depth by averaging the closest 40 light curve points immediately before and after the
transit to obtain a baseline.  We present an example of each transit in Figure~\ref{fig:candphase}.  

The properties derived are similar to other ground-based transit detections, but a detailed analysis,
including follow-up spectroscopic and imaging observations of these candidates is presented in  
Howell \etal (2010).          

\subsection{Two Specific Cases of Stellar Variability}
\label{sec:othervar}

During any large area photometric survey, there is the potential of
discovering unusual to rare objects.  In the case of BOKS, we detail here two
unusual variable stars that we have found in the survey.  

\subsubsection{BOKS-45906}
The first object, whose location is displayed in Figure~\ref{fig:dwarfnfc},
is known as BOKS-45906 (KIC 9778689).  On the first two clear nights of
our survey (MJD 3980 \& 3988), the star had a mean magnitude of $r \approx 20$,
though there were clear signs of variability of up to a magnitude in amplitude.
The star continued to vary on both nightly and intra-nightly timescales.  Then,
between MJD 4004 and 4005, the star had an eruption, reaching a maximum of 
$r=16.6 \pm 0.01$ on MJD 4006.  It  then declined in flux, returning to the approximate quiescent 
flux level on MJD 4009.  The overall light curve is plotted in Figure~\ref{fig:dwarfn}.  
From the light curve, it is likely that this star is a newly discovered 
cataclysmic variable of the dwarf nova subtype.  There is scientific interest in the light
curves of similar objects, both before and after eruption 
\citep{rob1975,collazzi2009,schaefer2010}: therefore additional photometric monitoring
of this source may be helpful. 

\subsubsection{BOKS-53856}
The second interesting variable we have found is a blue star in the
BOKS survey (a finding chart is displayed in Figure \ref{fig:bluefc}), known as 
BOKS-53856.  From comparison to the KIC78, it has a measured color of $g-r=-0.46$, making
it the bluest stellar source in our field.  Analysis of its light curve indicates
periodic variability, with a period of 0.255 days, though of an unusual nature.  The
phased light curve is presented in Figure \ref{fig:bluelc}.

We obtained a 900 second spectrum of BOKS-53856 using the Kitt Peak 2.1-m telescope 
and the GoldCam spectrograph on UT 26 June 2008.  We used the 300 l/mm grating (\#32) with a 
one arcsecond slit to provide a mean spectral resolution of 2.4\AA~across the full wavelength
range.  The spectra were reduced in the normal manner with observations
of calibration lamps and spectrophotometric standard stars obtained
before and after each sequence and bias and flat frames collected in 
the afternoon.  The final reduced spectrum is displayed in Figure 
\ref{fig:bluespec}.  

The most obvious features in the spectrum are the blue continuum and strong
Balmer lines, indicating a DA white dwarf type spectrum.  In general, 
blue variables are either of low amplitude and consist of pulsations, or as 
in the case here, show larger variations and may be some sort of interacting 
binary.  Holberg \& Howell (2011; in preparation) present a 
further study of this star.

\section{\label{sec:conclusions} Conclusions }

The goal of the BOKS survey was to constrain the amount and nature of
variability in a subsection of NASA's \kepler mission field of view.
The dedicated observations we conducted of the BOKS field allowed us
to observe variability on various timescales from a few minutes to
many days.  The long-term observations also allowed us a reasonable
opportunity to search for hot Jupiter exoplanet transits.

Through our preliminary analysis of the variability in the BOKS field 
we have identified $\sim$2,457 candidate variable stars with 776
candidate periodic variables.  Most of this periodic variation can be attributed
to rotating, low-mass stars with magnetic activity star/spots.  We have also found
over 90 $\delta$~Scuti stars, over 32 eclipsing binaries and contact
binaries, as well as tens of large amplitude pulsators, such as RR
Lyrae stars.  Within the BOKS field of view, we have also identified
at least three exoplanet candidates, all of which are undergoing 
observations by the \kepler mission.  The comparison of ground-based
and space-based transit observations should be beneficial to many future 
surveys.

\acknowledgements 

We would like to thank the entire staff of Case Western Reserve University
Warner \& Swassey observatory, including Heather L. Morrison, Charles Knox, 
and Colin Wallace for their invaluable assistance with the Burrell Schmidt.
We are also extremely grateful to the observers of the AAVSO who performed many 
observations of our field concurrently.  We thank Richard Wade for some useful
discussions.  We also thank the anonymous referee for several suggestions that
improved the quality of this paper.  This research has made use of the WEBDA database, 
operated at the Institute for Astronomy of the University of Vienna.  This publication 
makes use of data products from the Two Micron All Sky Survey, which is a joint project 
of the University of Massachusetts and the Infrared Processing and Analysis 
Center/California Institute of Technology, funded by the National Aeronautics 
and Space Administration and the National Science Foundation.

{\it Facilities:} \facility{KPNO:2.1m (Goldcam),} \facility{CWRU:Schmidt}

\pagebreak

\pagebreak
\begin{deluxetable}{lccccl}
\tablewidth{0pt}
\tablenum{1}
\tabletypesize{\footnotesize}
\tablecaption{Observing Log Summary\label{table:obslog}}
\tablehead{
\colhead{Night}
& \colhead{JD}
& \colhead{Hours Used}
& \colhead{N$_{ images}$}
& \colhead{Airmass Range}
& \colhead{Notes}
}
\startdata
1  & $2453980$  &  1.68  &  25 & 1.06 -- 1.24 &                        \\ 
2  & $2453981$  &  0.00  &  -- & --           & unusable due to weather\\ 
3  & $2453982$  &  0.00  &  -- & --           & unusable due to weather\\ 
4  & $2453983$  &  0.00  &  -- & --           & unusable due to weather\\ 
5  & $2453984$  &  0.00  &  -- & --           & unusable due to weather\\ 
6  & $2453985$  &  0.00  &  -- & --           & unusable due to weather\\ 
7  & $2453986$  &  0.00  &  -- & --           & unusable due to weather\\ 
8  & $2453987$  &  0.00  &  -- & --           & unusable due to weather\\ 
9  & $2453988$  &  0.42  &   6 & 1.19 -- 1.25 & \\ 
10 & $2453989$  &  4.68  &  56 & 1.03 -- 1.71 & \\ 
11 & $2453990$  &  0.00  &  -- & --           & unusable due to weather\\ 
12 & $2453991$  &  0.00  &  -- & --           & removed - CCD condensation \\ 
13 & $2453992$  &  0.00  &  -- & --           & removed - variable clouds\\ 
14 & $2453993$  &  5.96  &  69 & 1.04 -- 2.16 & \\ 
15 & $2453994$  &  5.87  &  81 & 1.04 -- 2.08 & \\ 
16 & $2453995$  &  6.56  &  95 & 1.04 -- 2.51 & \\ 
17 & $2453996$  &  6.33  &  92 & 1.03 -- 2.82 & \\ 
18 & $2453997$  &  6.95  &  98 & 1.04 -- 3.01 & \\ 
19 & $2453998$  &  0.00  &  -- & --           & removed - CCD electronics\\ 
20 & $2453999$  &  2.63  &  38 & 1.04 -- 1.13 & \\ 
21 & $2454000$  &  1.79  &  24 & 1.04 -- 1.05 & \\ 
22 & $2454001$  &  6.54  &  86 & 1.04 -- 2.55 & \\ 
23 & $2454002$  &  0.00  &  -- & --           & removed - variable clouds\\ 
24 & $2454003$  &  6.64  &  91 & 1.04 -- 2.77 & \\ 
25 & $2454004$  &  6.73  &  93 & 1.04 -- 2.70 & \\ 
26 & $2454005$  &  6.60  &  93 & 1.04 -- 2.72 & \\ 
27 & $2454006$  &  6.68  &  96 & 1.04 -- 2.84 & \\ 
28 & $2454007$  &  5.86  &  84 & 1.04 -- 2.08 & \\ 
29 & $2454008$  &  6.25  &  90 & 1.04 -- 2.45 & \\ 
30 & $2454009$  &  4.75  &  66 & 1.04 -- 1.58 & \\ 
31 & $2454010$  &  5.76  &  78 & 1.03 -- 2.18 & \\ 
32 & $2454011$  &  5.13  &  57 & 1.03 -- 1.97 & \\ 
33 & $2454012$  &  5.28  &  73 & 1.03 -- 2.21 & \\ 
34 & $2454013$  &  0.00  &  -- & --           & unusable due to weather \\ 
35 & $2454014$  &  0.00  &  -- & --           & unusable due to weather \\ 
36 & $2454015$  &  5.61  &  74 & 1.03 -- 2.59 & \\ 
37 & $2454016$  &  0.00  &  -- &              & removed - lunar background\\ 
38 & $2454017$  &  0.00  &  -- &              & unusable due to weather \\ 
39 & $2454018$  &  0.00  &  -- &              & unusable due to weather \\ 
40 & $2454019$  &  0.00  &  -- &              & unusable due to weather \\ 
\enddata
\end{deluxetable}
\vfill
\pagebreak

\begin{deluxetable}{lcccccccl}
\tablewidth{0pt}
\tablenum{4}
\tabletypesize{\footnotesize}
\tablecaption{BOKS Exoplanet Candidates\label{table:bokscand}}
\tablehead{
\colhead{BOKS}
& \colhead{KIC}
& \colhead{RA}
& \colhead{Dec}
& \colhead{$r$}
& \colhead{$g-r$\tablenotemark{a}}
& \colhead{Period}
& \colhead{Eclipse Depth}
& \colhead{Notes}\\
\colhead{ID}
& \colhead{ID}
& \colhead{(J2000)}
& \colhead{(J2000)}
& \colhead{(mag)}
& \colhead{(mag)}
& \colhead{(days)}
& \colhead{(mag)}
& \colhead{}
}
\startdata
40959 & 9595827 & 19 39 27.667 & 46 17 09.23 & 15.1 & 0.63 & 3.9 & 0.02 $\pm$ 0.01 & \\
45069 & 9838975 & 19 40 08.003 & 46 36 01.22 & 16.1 & 0.73 & 2.6 & 0.04 $\pm$ 0.01 & \\
52481 & 9597095 & 19 41 18.802 & 46 16 06.00 & 15.9 & 0.63 & 7   & 0.05 $\pm$ 0.01 & Period approximate\\
\enddata
\tablenotetext{a}{For reference, the Sun is believed to have a $g-r$ color of 
0.44 $\pm$ 0.02 \citep{bilir2005,rodgers2006}} 
\end{deluxetable}

\begin{figure}
\figurenum{2}
\caption[ BOKS field ]{\label{fig:field} An image of the
BOKS field, created by combining the first 24 images of the
survey.  This image is approximately $101\parcmin5$ by $49\parcmin5$ in size.  
North is up and East is to the left.  The open cluster NGC~6811,
whose center is located at $\alpha = 19^{h}37^{m}17^{s}$, $\delta = +46^{d}23^{m}18^{s}$ 
(WEBDA database) is clearly visible in the center of the image.}
\end{figure}

\begin{figure}
\figurenum{3}
\plotone{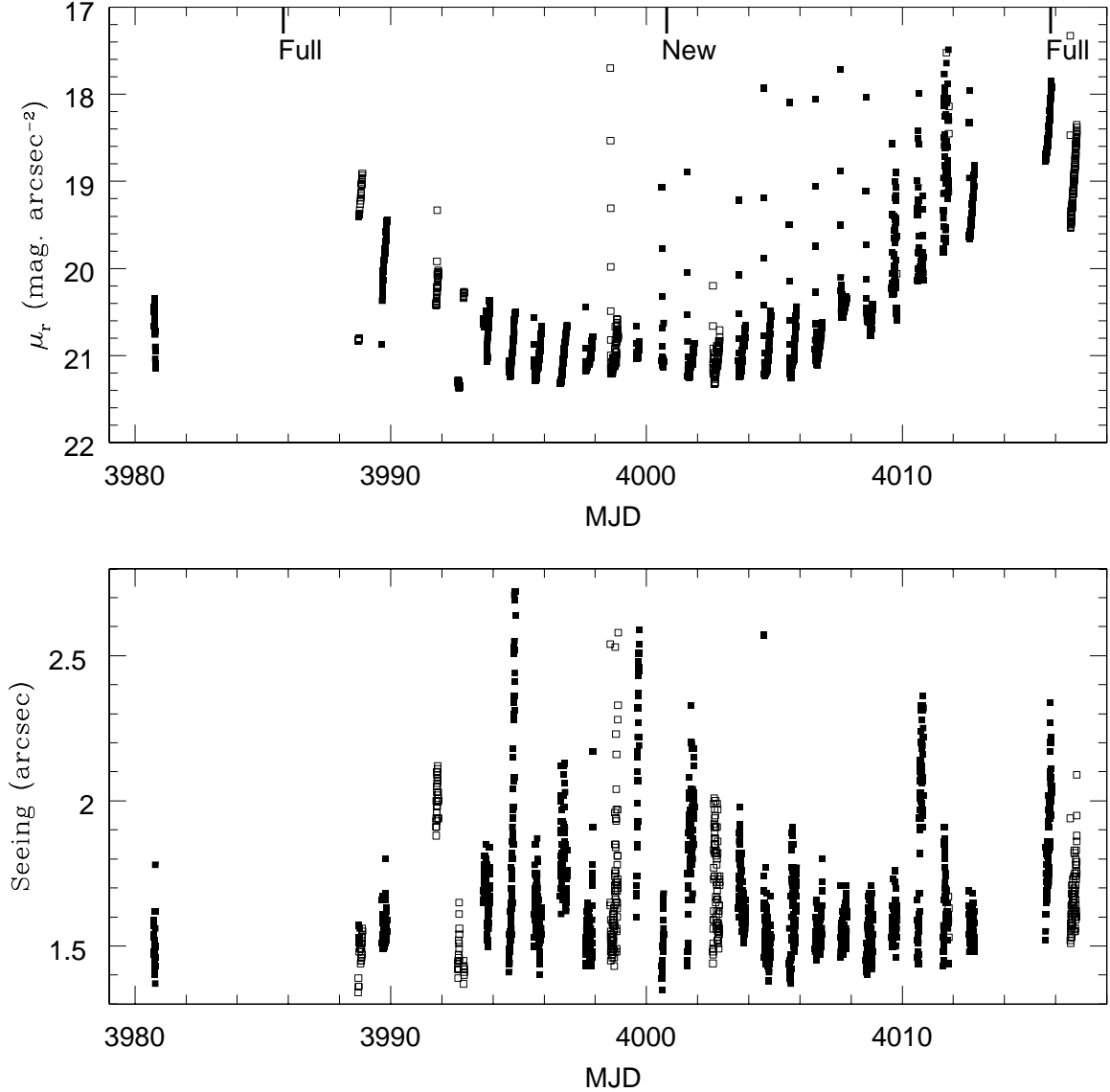}
\caption[Sky Transparency]{\label{fig:skysee} The median seeing and median
sky brightness for all of the $r$ images in the BOKS survey.  Filled squares denote
the points used in the final light curve and open squares denote omitted points.  Given
the long length of the survey (40 nights), there are broad differences in these parameters.  Note that
the derived surface brightness comes from lunar phase differences, the presence and
absence of clouds, and also the effect of astronomical twilight.  The times of the full and new
moon are shown for reference.  The seeing values are measured from the images themselves, 
which have a pixel scale of 1.45\arcsec, and are likely overestimates of the ambient seeing 
at the time of the observations.}
\end{figure}

\begin{figure}
\figurenum{4}
\plotone{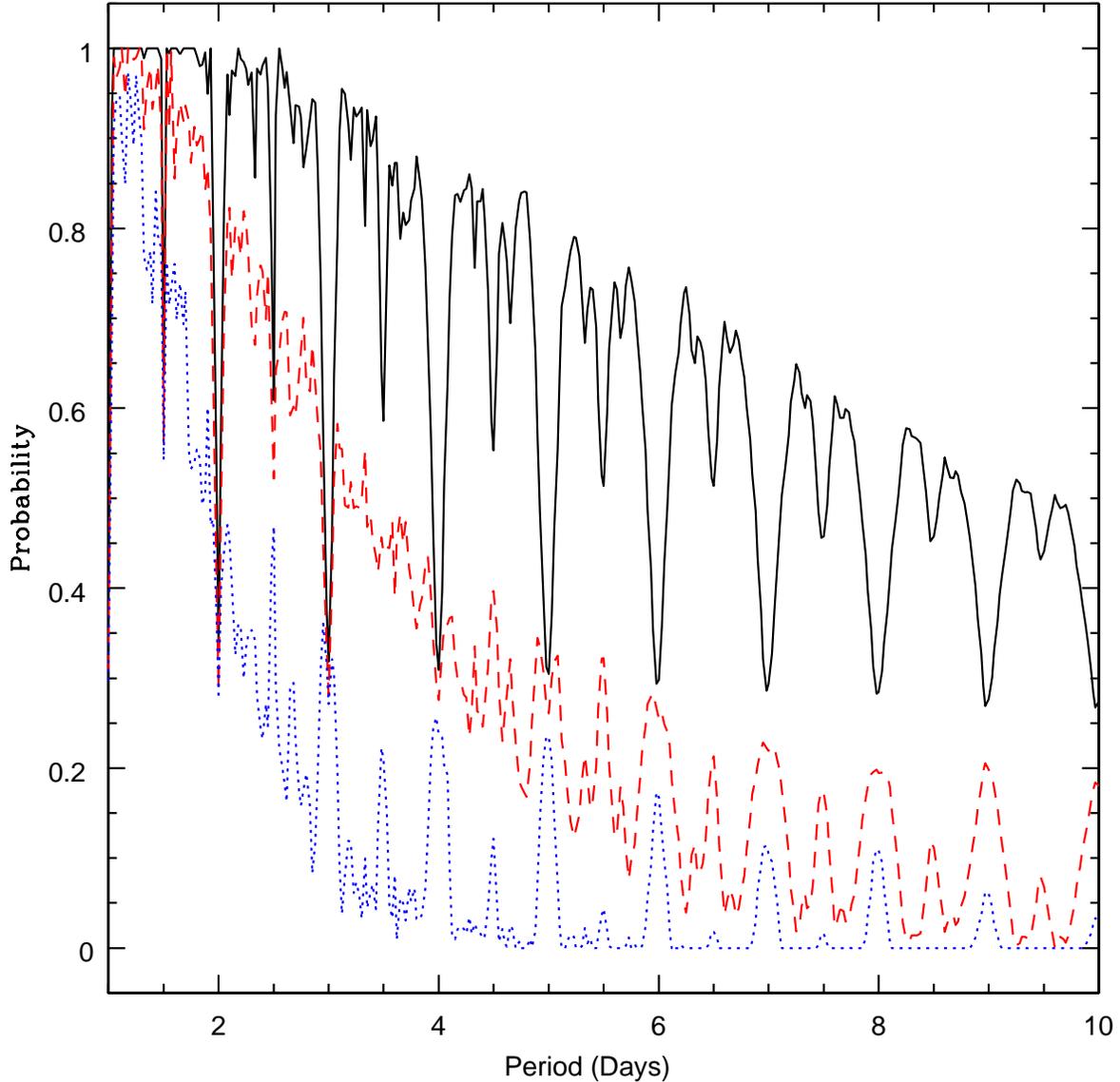}
\caption[ Window function for BOKS ]{\label{fig:window} The temporal window
function for the BOKS data set.  The solid black line indicates the likelihood
of viewing one transit event, the dashed red line indicates the likelihood
of viewing two transit events, and the dotted blue line indicates the
likelihood of viewing three events, as a function of the system's period.
Given the ground-based nature of BOKS, there are strong aliasing effects
at periods of integer days.  Note that this window function is approximate: 
see \S 3 for discussion.}
\end{figure}

\begin{figure}
\figurenum{5}
\plotone{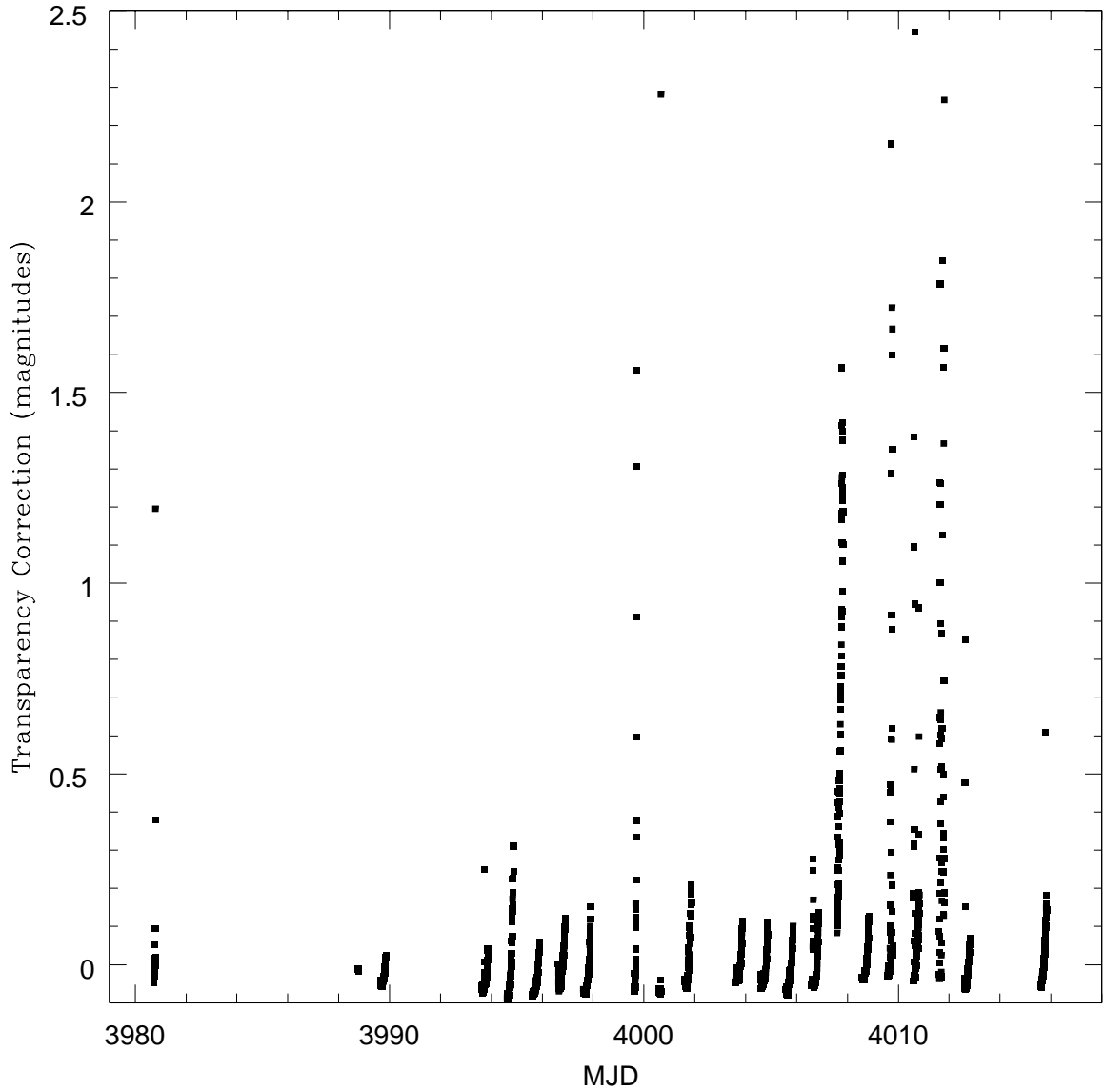}
\caption[Transparency]{\label{fig:offsets}A plot of the magnitude correction derived from 
the ensemble stars for a single region over the entire BOKS survey period.  Note that this correction 
takes into account the effects of airmass, seeing, and transparency simultaneously.  The airmass effect of 
each night's observations is clearly visible in the data, as the BOKS field was close to zenith at 
the beginning of each night, and set as the night progressed.}
\end{figure}

\begin{figure}
\figurenum{6}
\plotone{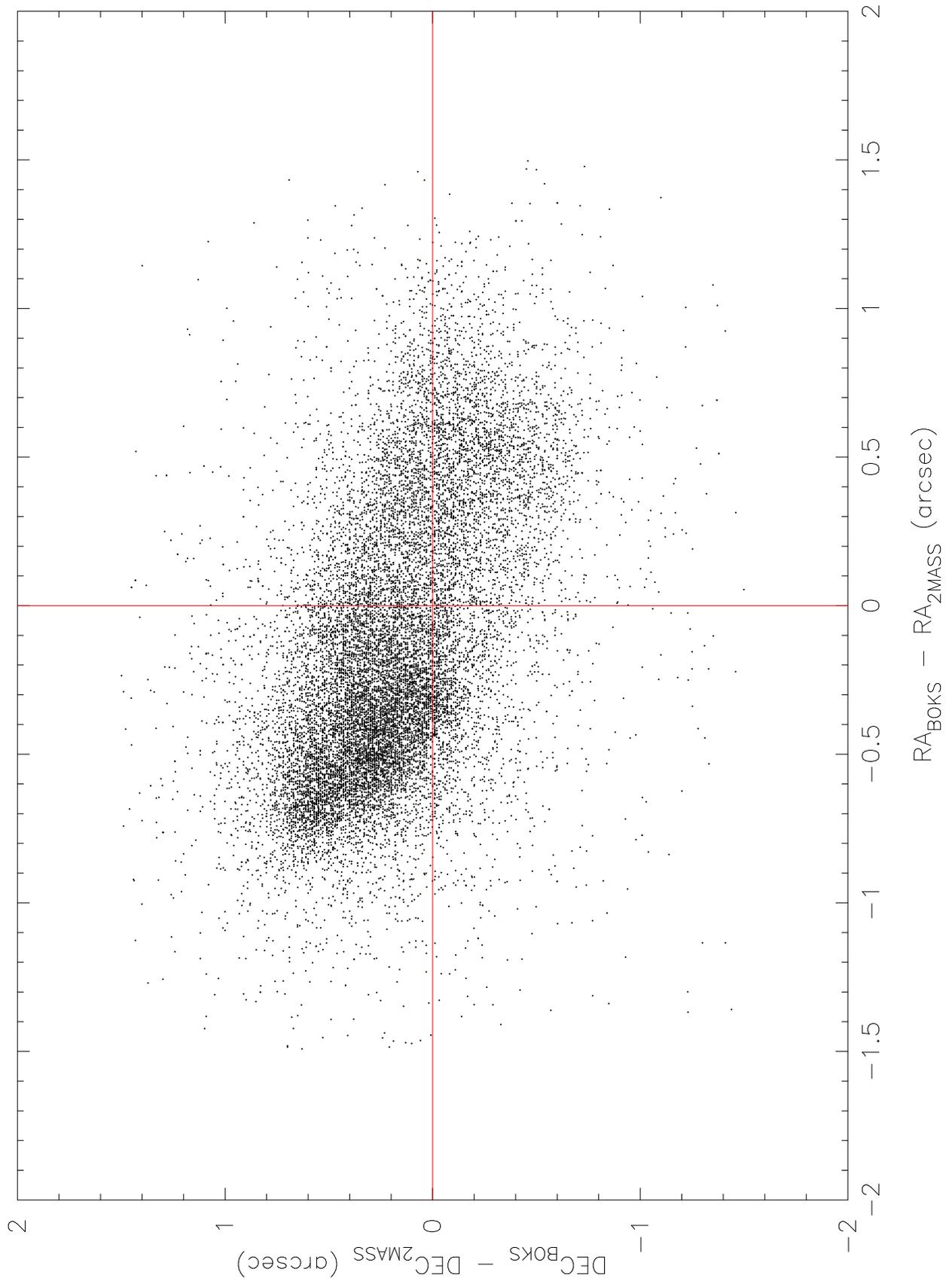}
\caption[ Short ]{\label{fig:match} The deviation of the BOKS astrometric coordinates to 
the coordinates found from the 2MASS survey.  These deviations should be compared to the
pixel scale of the Burrell, which is 1.45\arcsec.  }
\end{figure}

\begin{figure}
\figurenum{7}
\plotone{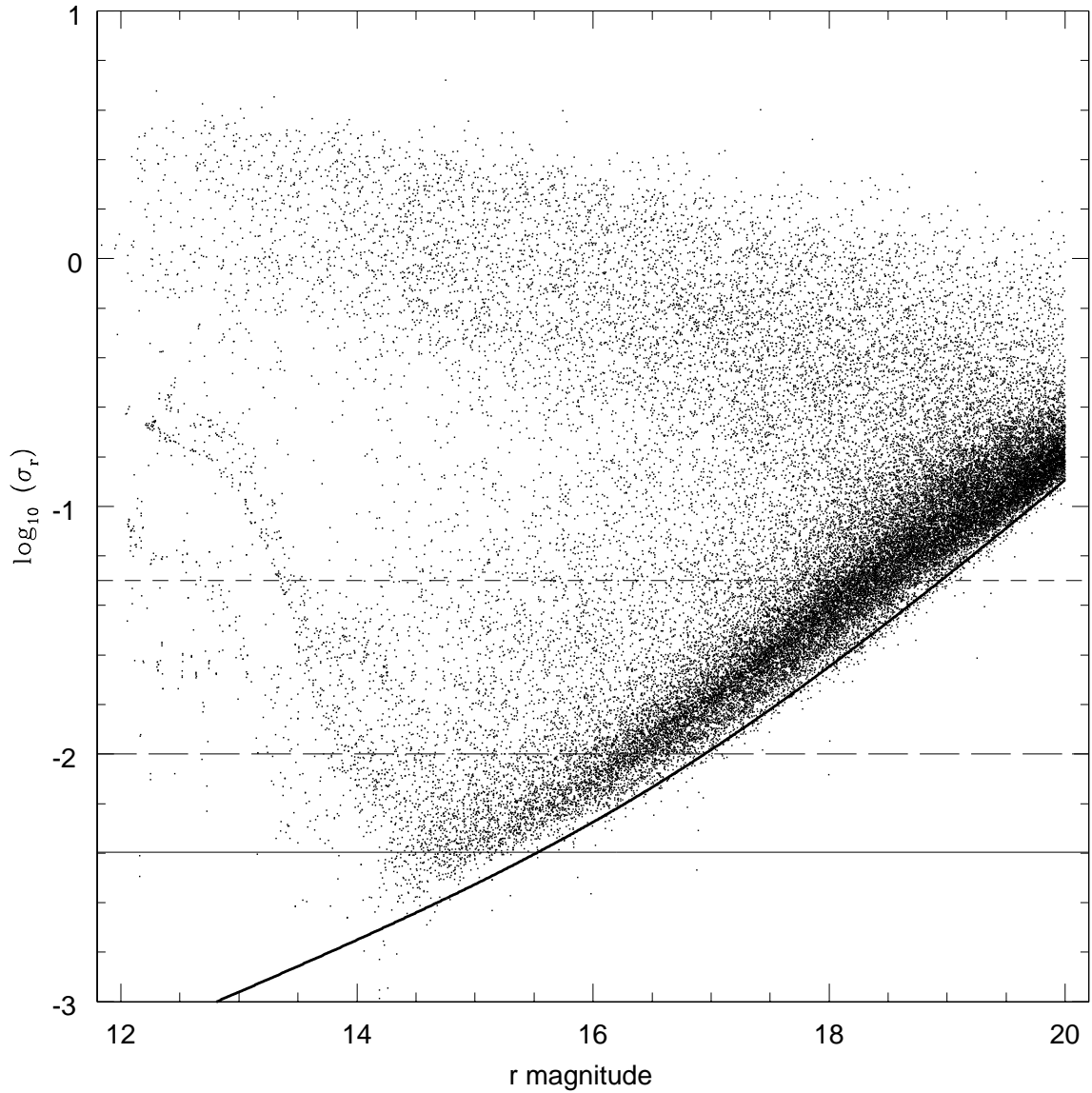}
\caption[ Error of each star ]{\label{fig:sigma} The standard
deviation of the light curve of each star as a function of $r$
magnitude, as determined by OPTICSTAT, for all stars brighter
than an $r$ magnitude of 20.  Note that this plot includes every source,
so variable stars will lie above the diagonal sequence that denotes the
photometric error function. The upward bending of the points brighter than $r$ = 14. 
is due to saturation effects.  The thick diagonal line denotes the expected photometric
uncertainty for a source observed at new moon, and should denote the lower edge of the
true uncertainty distribution.  The solid horizontal line shows a standard deviation of 4mmag, the
long horizontal line shows a standard deviation of 10mmag and the short dashed horizontal line 
shows a standard deviation of 50 mmag, for reference.}
\end{figure}

\begin{figure}
\figurenum{8}
\plotone{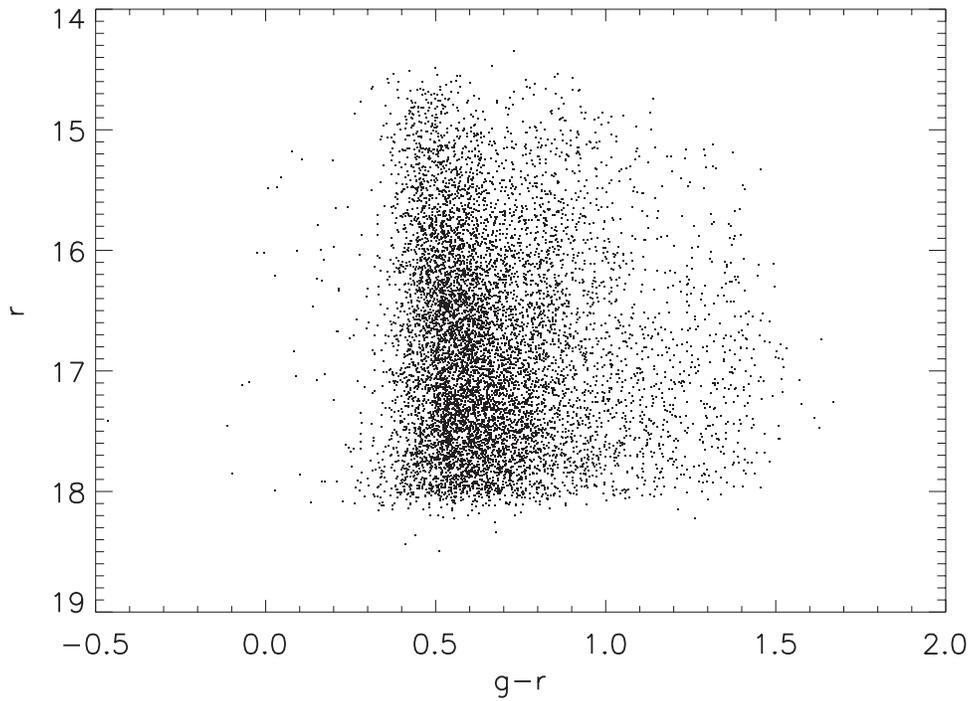}
\caption[CMD of Constant Stars]{\label{fig:const_grr_cmd} A g$-$r CMD
for the 10,881 non-variable stars detected with this survey that have entries in the Kepler
Input Catalog.  Note the single broad sequence of stars at $g-r \approx 0.5$, which corresponds to
stars at the main-sequence turn-off for an old stellar population.  This CMD 
can be compared to the much deeper data of \citet{dejong2010}, and shows that
the BOKS field is typical of fields observed through the Milky Way's disk.}
\end{figure}

\begin{figure}
\figurenum{9}
\plotone{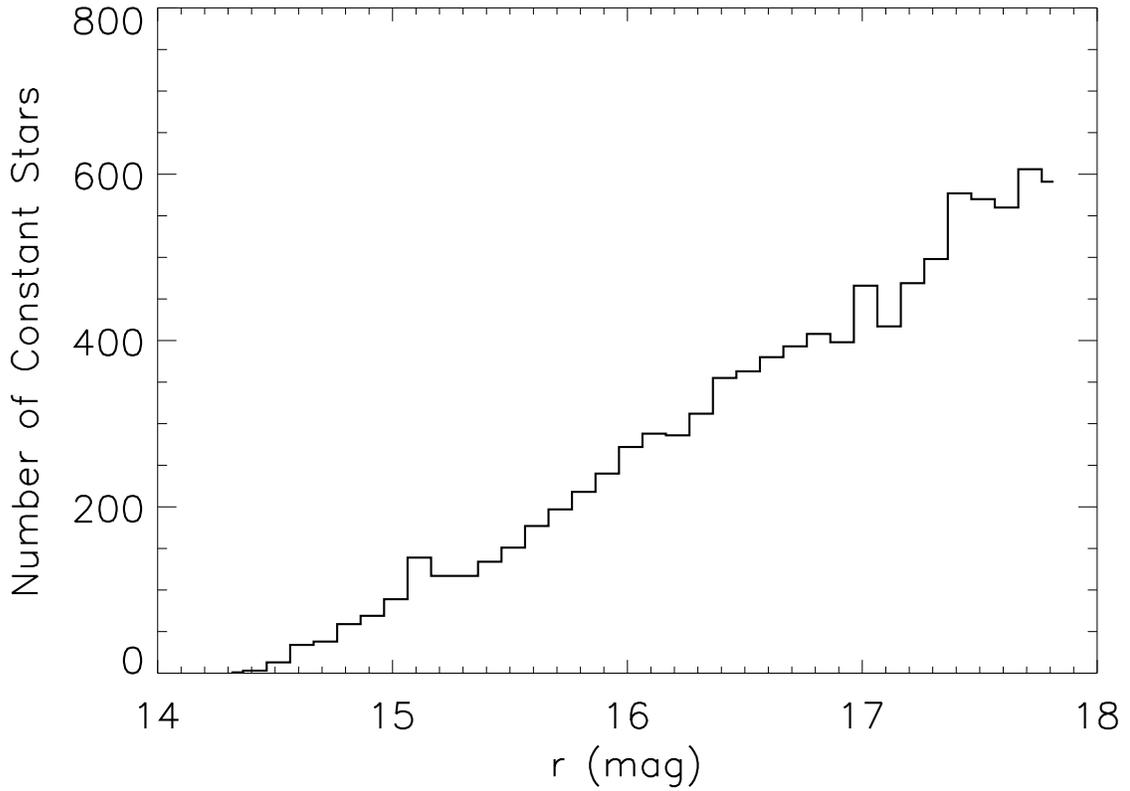}
\caption[Number of Constant Stars vs. r
magnitude]{\label{fig:const_n_vs_r} A comparison of the number of photometrically
constant stars as a function of magnitude.  As expected, the number of photometrically
constant stars rises with apparent magnitude since the total number of stars increases, 
and our ability to search for photometric variations depends strongly on signal-to-noise.}
\end{figure}

\clearpage
\begin{figure}
\figurenum{10}
\plotone{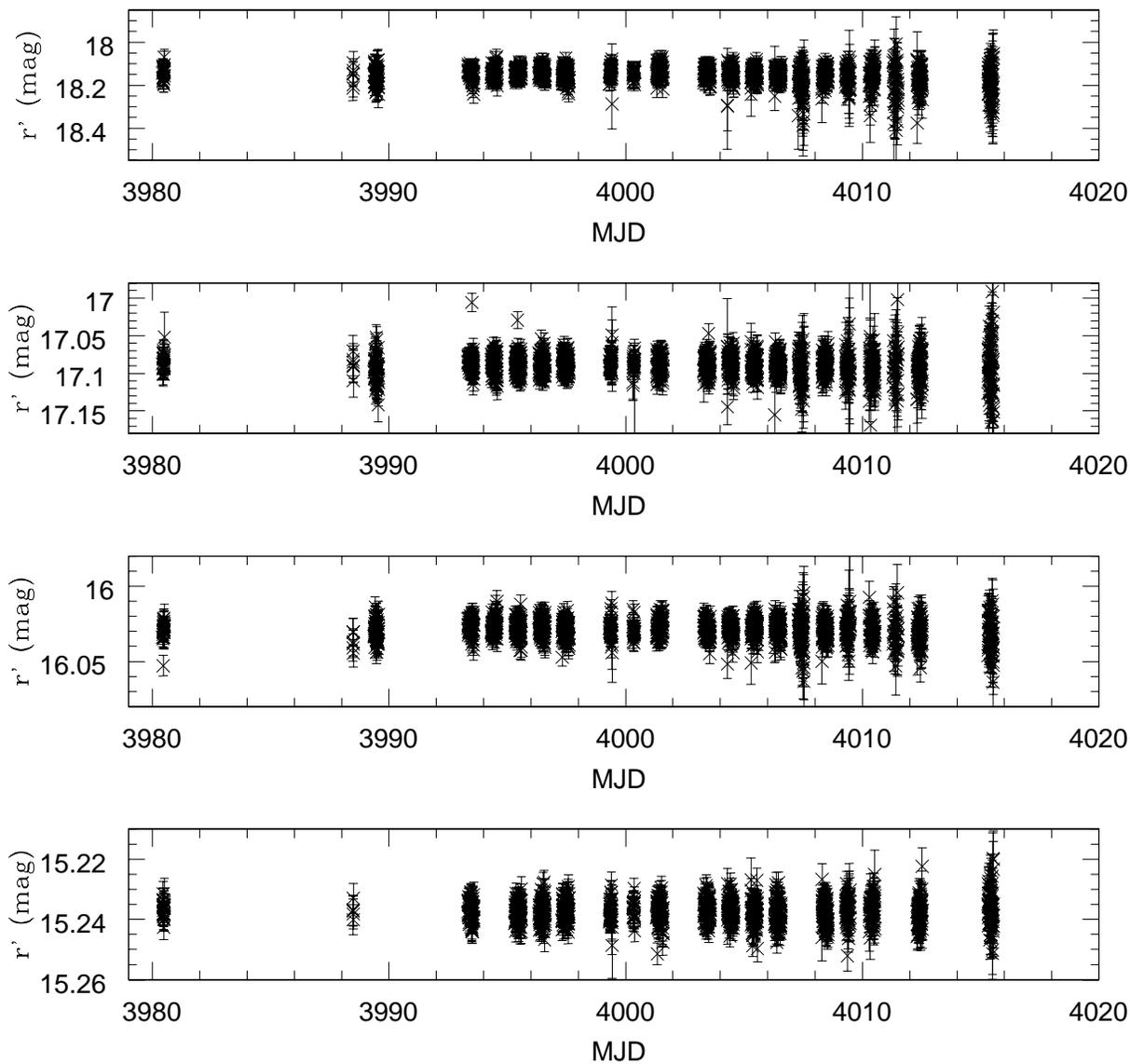}
\caption[Constant Star ]{\label{fig:constar} An example of four stars with differing
magnitudes that were found to be 
photometrically constant throughout the survey, using the criteria for variability.
From top to bottom the stars are BOKS-33934, BOKS-11505, BOKS-4700, and BOKS-3767.
Note that the magnitude axis is substantially different for each star, ranging from
700 mmag for BOKS-33934 down to 50 mmag for BOKS-3767.  Many of the most deviant points 
in each of the light curves are due to radiation events (cosmic rays) hitting the star's aperture.}
\end{figure}

\begin{figure}
\figurenum{11}
\plotone{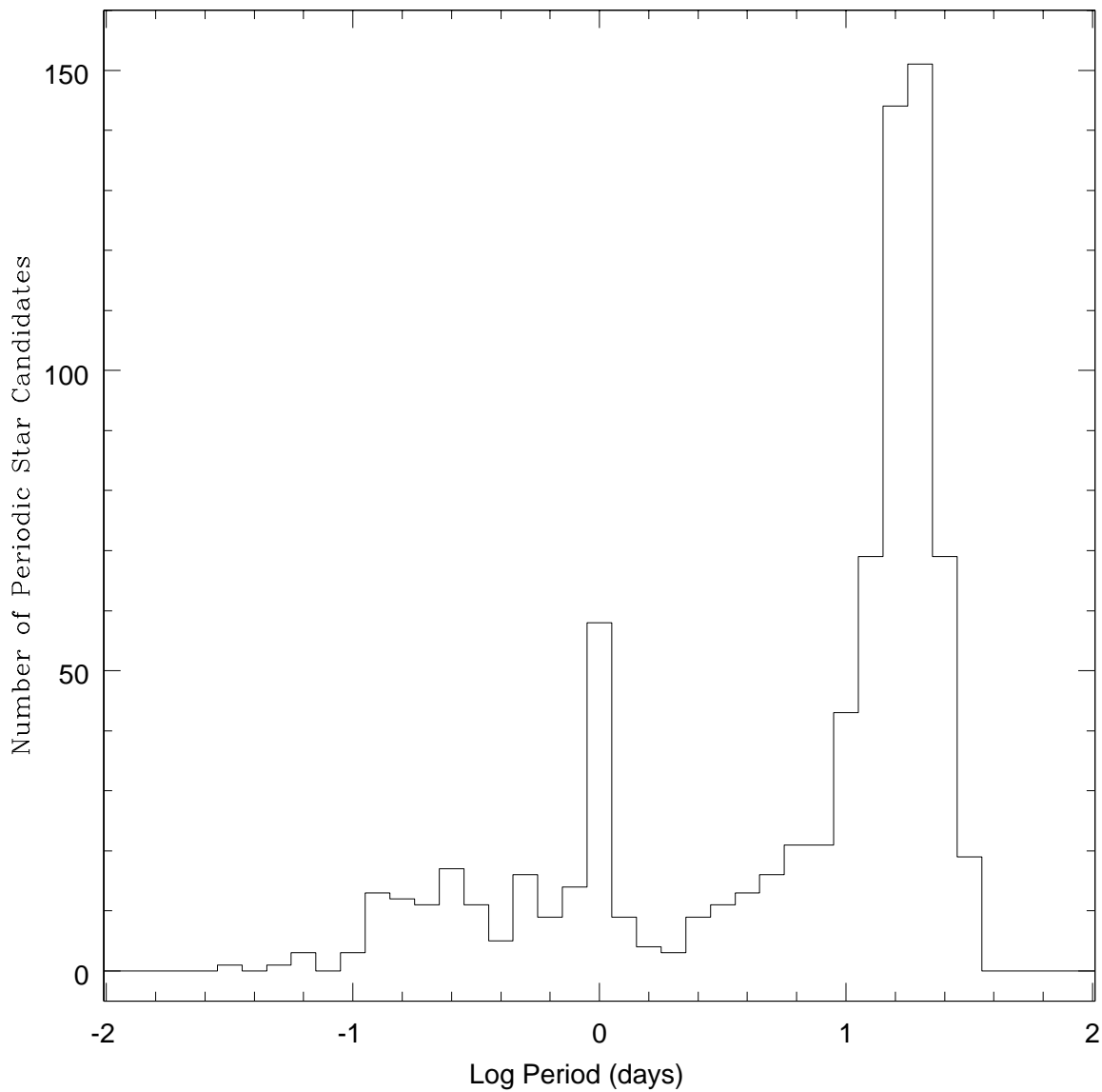}
\caption[ Maximum Amplitude of the stars ]{\label{fig:logp} A
histogram of the number of periodic stars in our sample with r
magnitudes brighter than 18.5 versus the logarithm of the derived 
period. Roughly half of the periodic stars we have identified have periods 
longer than 10 days.  There is also a noticeable spike in the distribution near 1 day, suggesting
that some of the periods we have identified are due to aliasing near the cadence of our observations, 
despite the fact we removed objects with periods of 1 $\pm$ 0.025 days (\S 5.2.2).}
\end{figure}

\begin{figure}
\figurenum{12}
\plotone{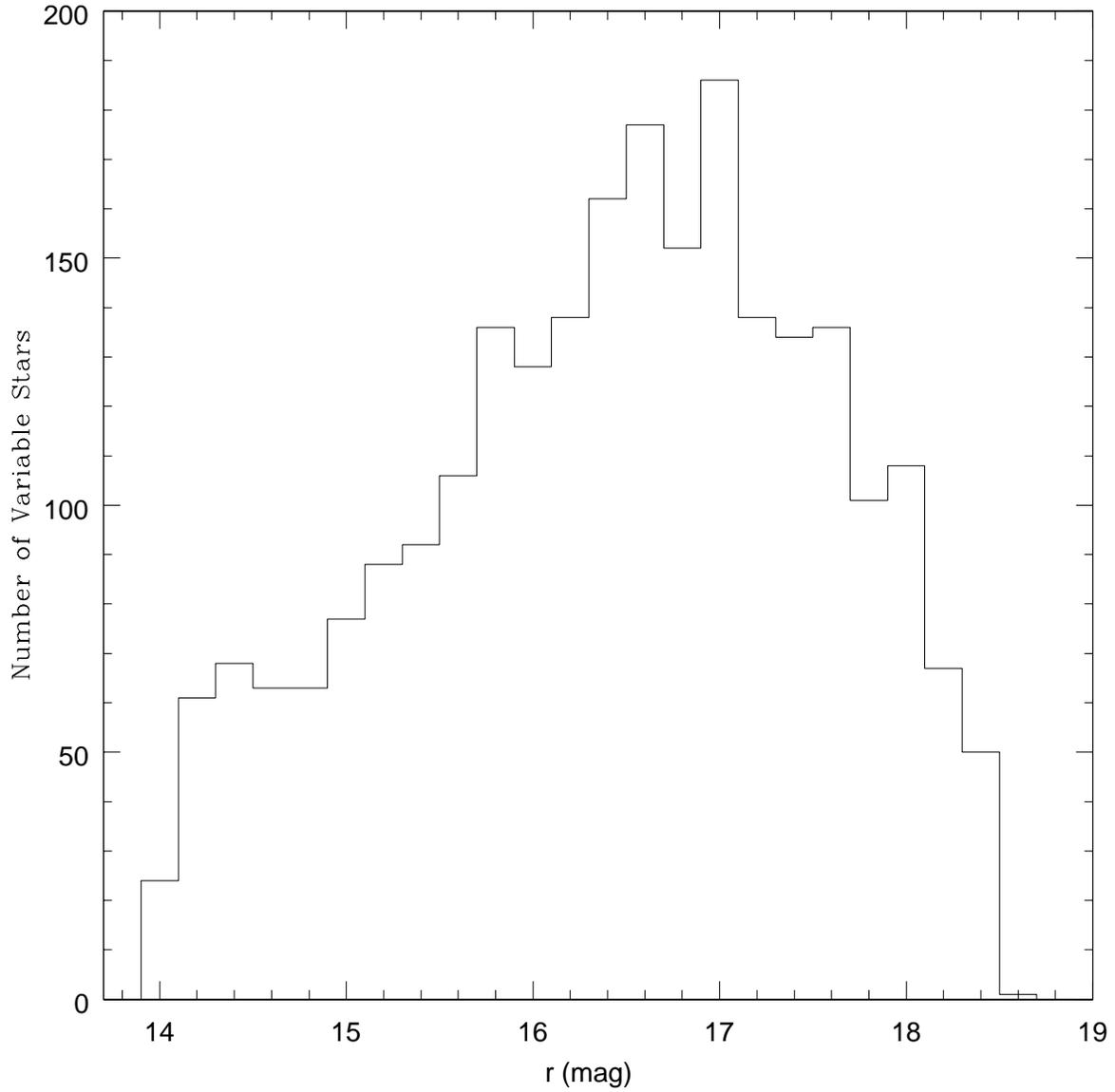}
\caption[Histogram of Variable Stars vs r
magnitude]{\label{fig:var_n_vs_r} A histogram of the number of
periodic variable stars as a function of r magnitude.  The number of 
variable stars detected rises with increasing magnitude down to r$\sim$16.  The 
number of detected variables drops rapidly between r$\sim$17 and r$\sim$19.  This
reflects the loss of sensitivity to low amplitude variables, such as
stars with rotational modulation, as the photometric uncertainties
increase.}
\end{figure}

\begin{figure}
\figurenum{13}
\plotone{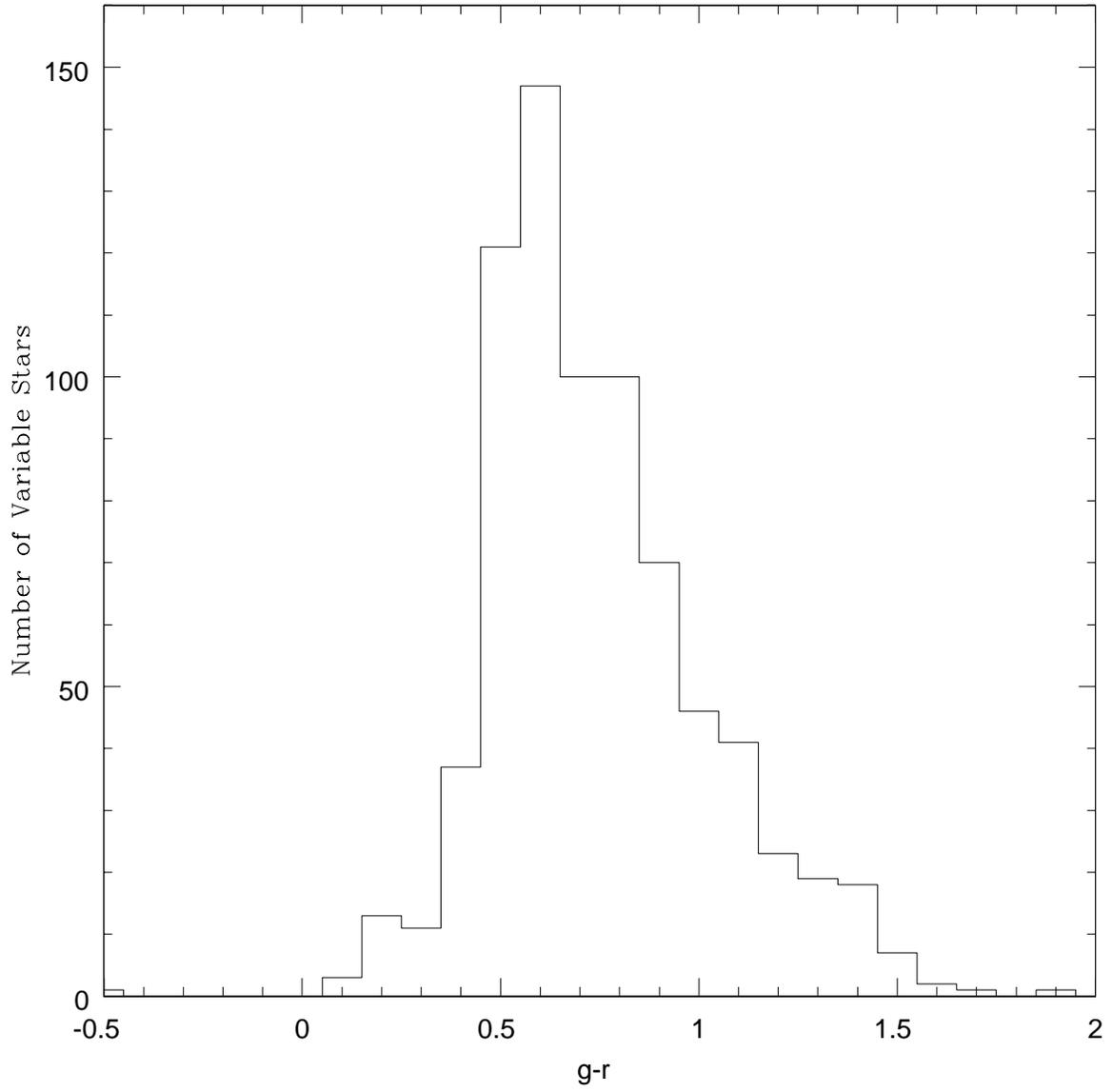}
\caption[Number of variable Stars vs
rmagnitude]{\label{fig:var_n_vs_gr} A histogram of the number of
periodic variables with r brighter than 18.5 vs g$-$r color.  This
distribution has a single peak near g$-$r=0.6 and a tail that extends
to g$-$r$\sim$1.6. This is similar to the color-magnitude diagram
of non-variable stars (Figure~\ref{fig:const_grr_cmd}),
except that there may be a small excess of variables near g$-$r=0.2.}
\end{figure}

\begin{figure}
\figurenum{14}
\plotone{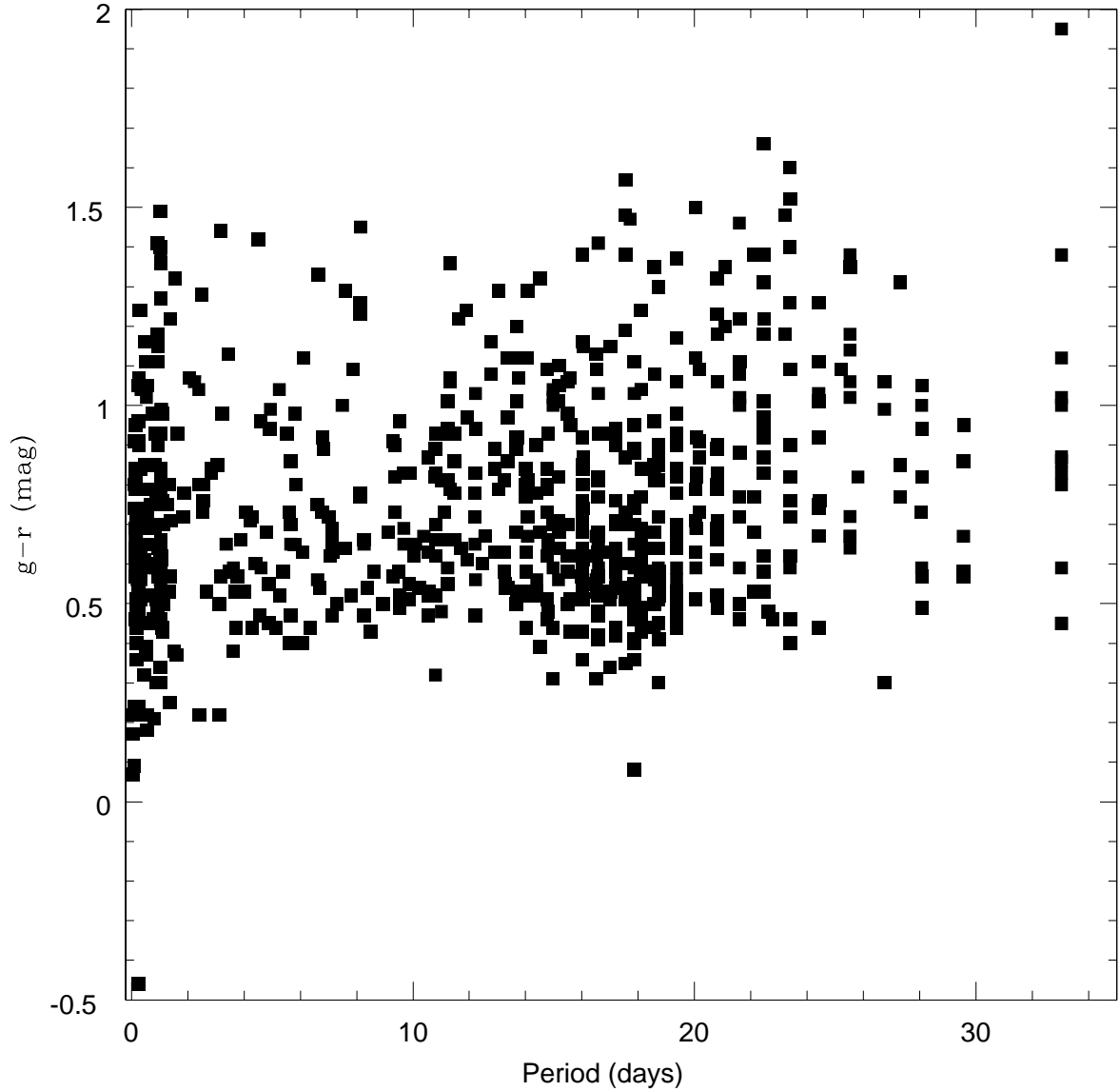}
\caption[Period vs g$-$r color]{\label{fig:var_p_vs_gr} A plot of
the period of the periodic variable stars versus their g$-$r color.
As can be clearly seen, most variables have colors that are similar to those of photometrically 
constant stars at all periods.  For the shortest period stars, P$\lesssim$1 day, 
there seems to be a small excess of stars with 0$\leq$g$-$r$\leq$0.5. }
\end{figure}

\begin{figure}
\figurenum{15}
\plotone{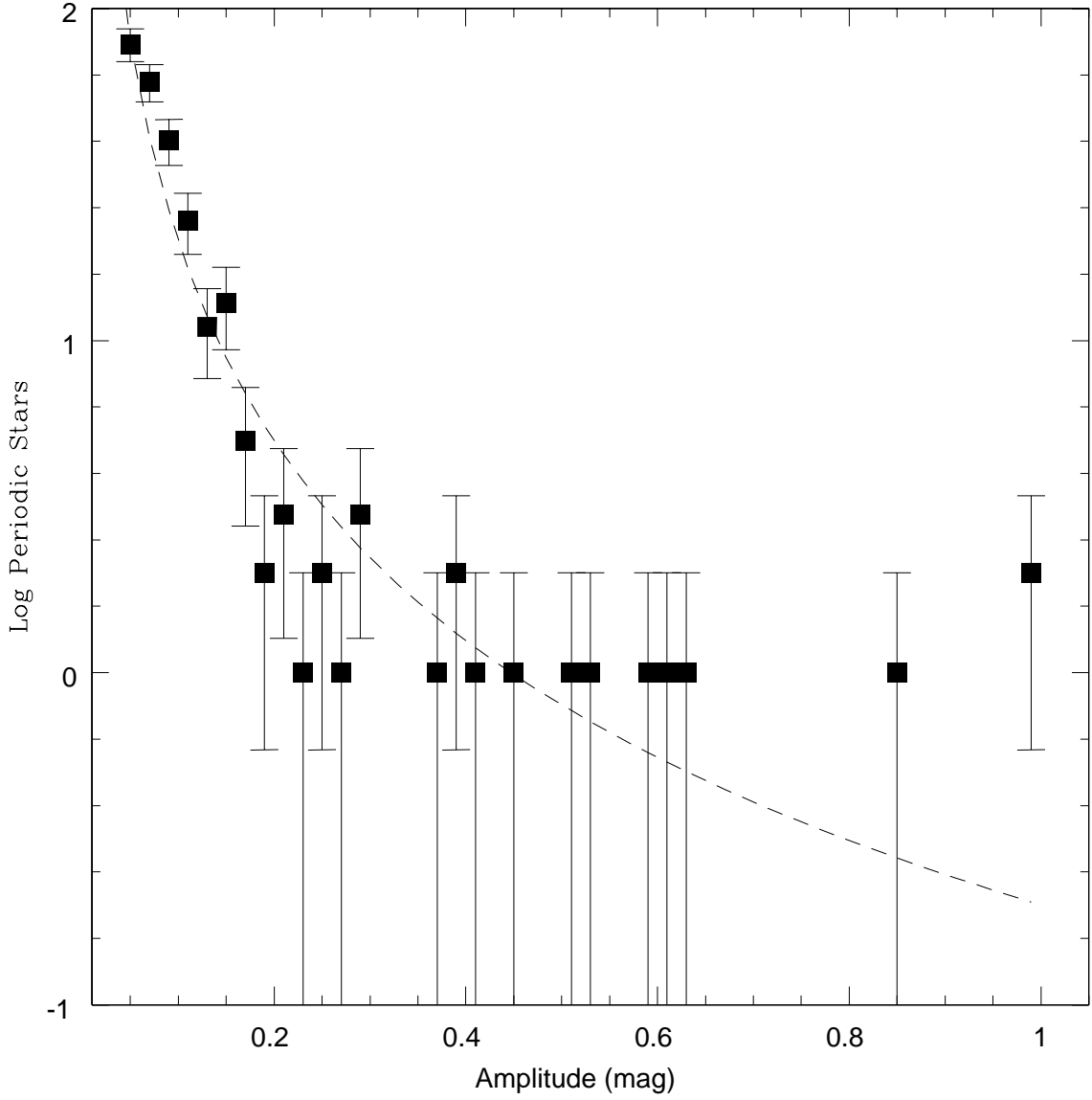}
\caption[Amplitude amount]{\label{fig:amplitude} The distribution of
periodic variables found in BOKS, as a function of photometric amplitude.  
The dashed line is the predicted function of variability from \citet{tonry2005}.  
The error bars are the Poisson uncertainties for each bin.}
\end{figure}

\begin{figure}
\figurenum{16}
\plotone{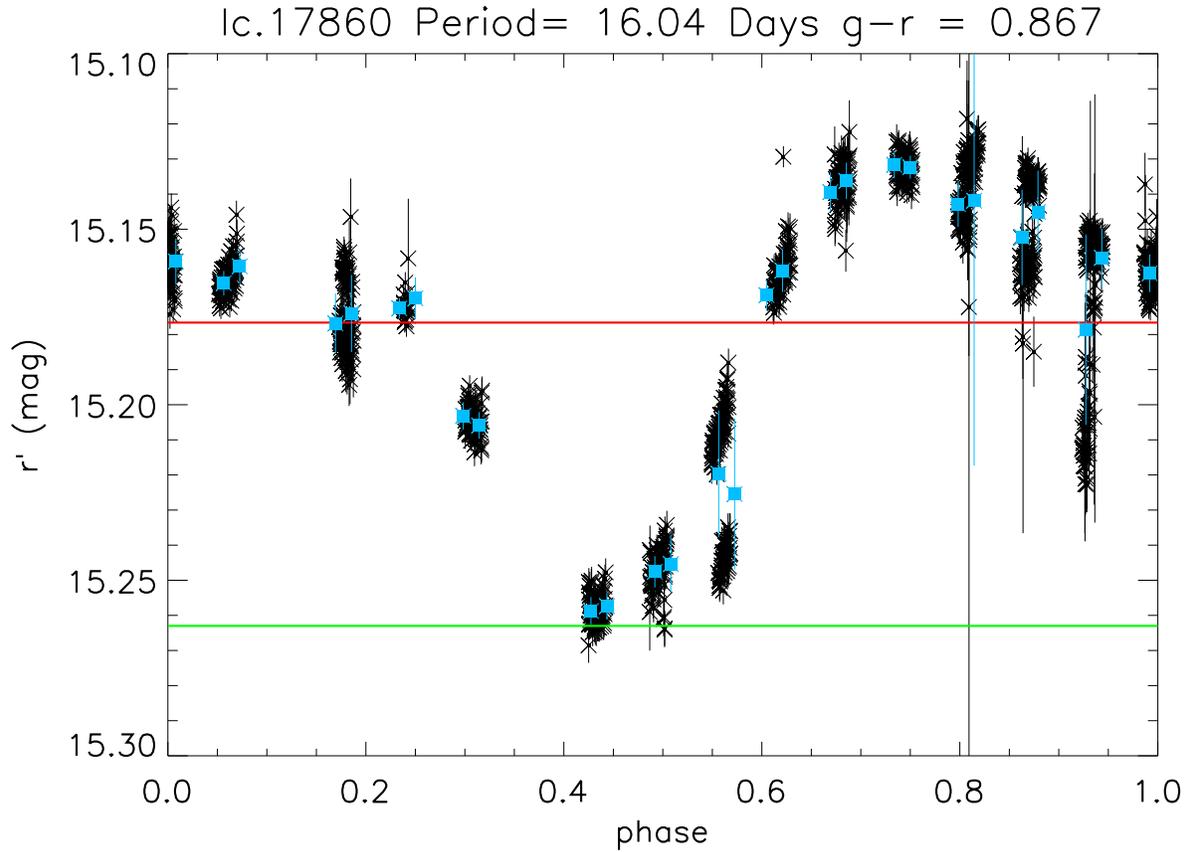}
\caption[ Pulsating variable ]{\label{fig:pulsvar} This phased light
curve is an example of a pulsating variable star from our survey.  The
black points are the individual photometric measurements (and error
bars).  The blue squares are weighted averages calculated for every 26
points of the phased light curve.  The red horizontal line at $r \approx 15.18$ is the
weighted average $r$ magnitude of the entire light curve.  The green horizontal
line at $r \approx 15.26$ is the magnitude listed for the star in the Kepler Input Catalog (KIC).}
\end{figure}

\begin{figure}
\figurenum{17}
\plotone{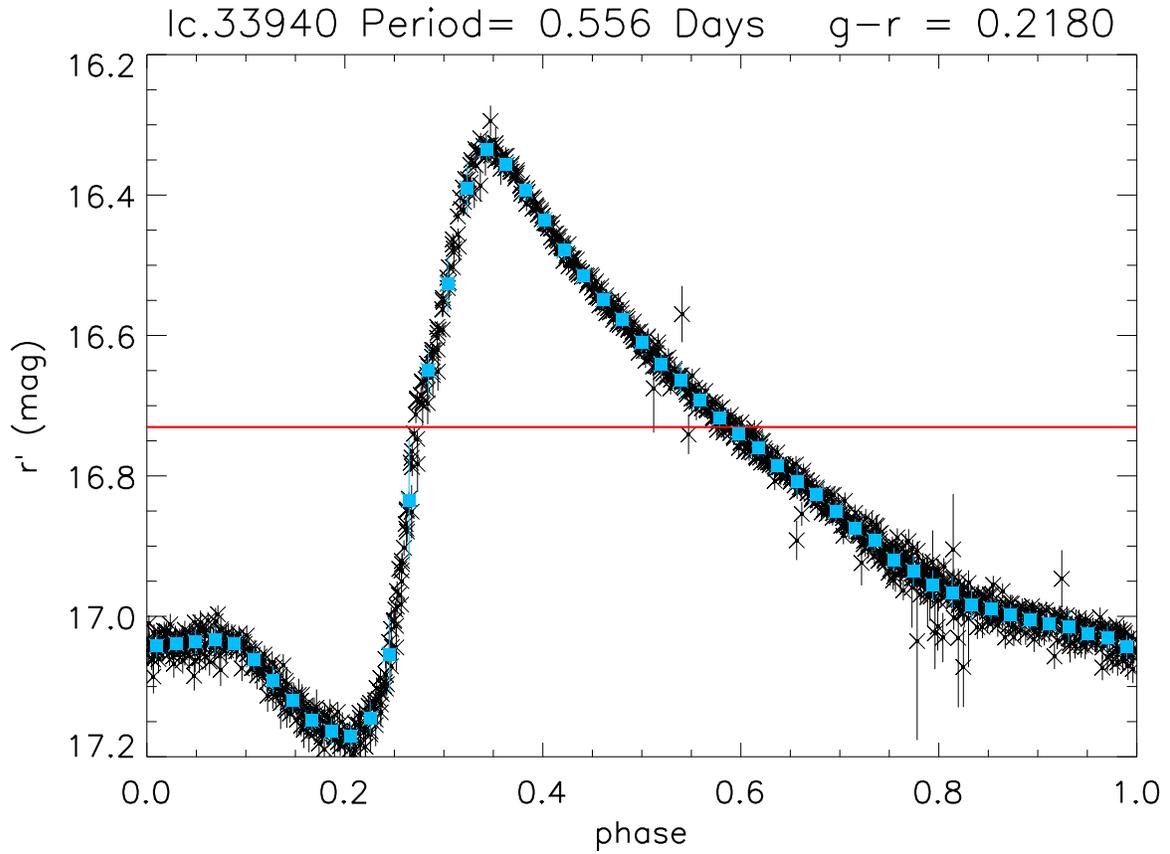}
\caption[ RR Lyrae ]{\label{fig:rrlyrae} This phased light curve is one example of 
a RR Lyrae star found within the BOKS field.  From visual inspection, the star
appears to be of the RRab subtype.  The symbols in this plot have identical properties
to those found in Figure~\ref{fig:pulsvar}, except that only the mean magnitude is
displayed as a horizontal line.}
\end{figure}

\begin{figure}
\figurenum{18}
\plotone{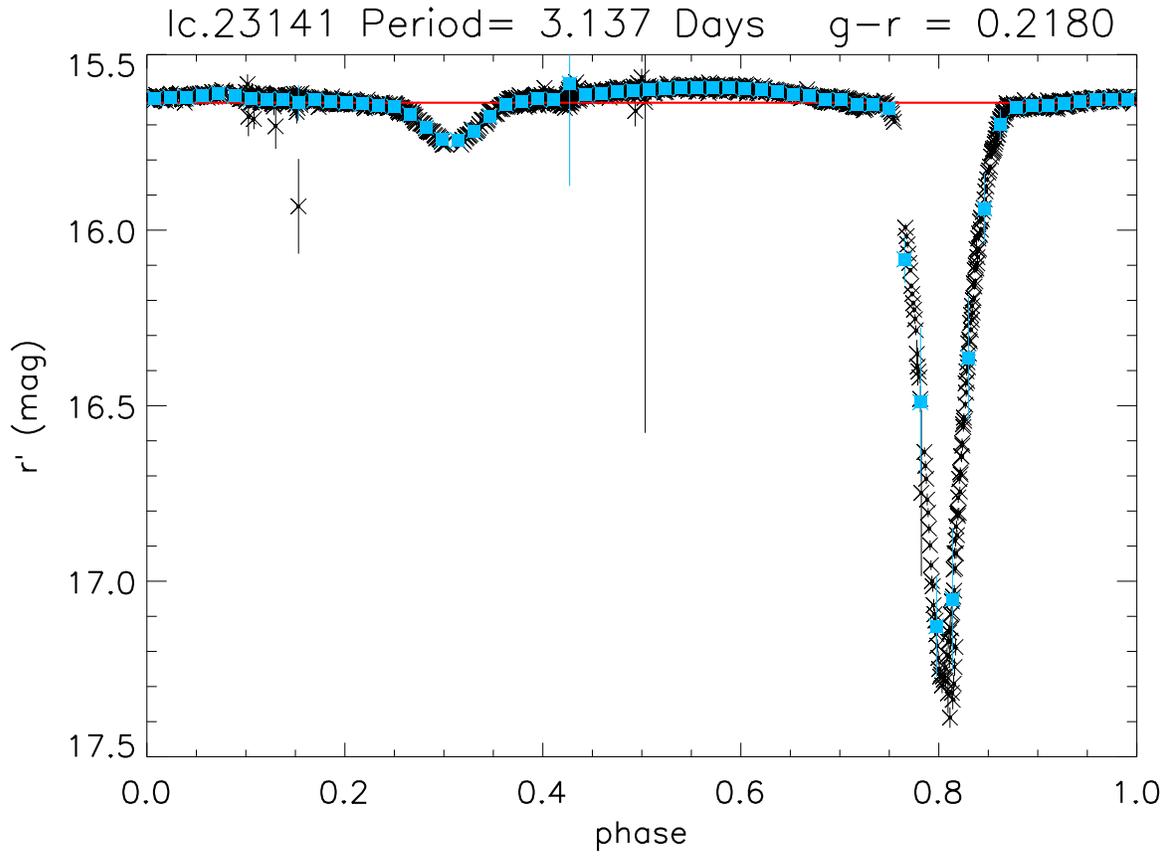}
\caption[ Eclipsing binary ]{\label{fig:eb1} This phased light curve
is an example of an eclipsing binary of the Algol type. The symbols are 
the same as in Figure~\ref{fig:rrlyrae}.}
\end{figure}

\begin{figure}
\figurenum{19}
\plotone{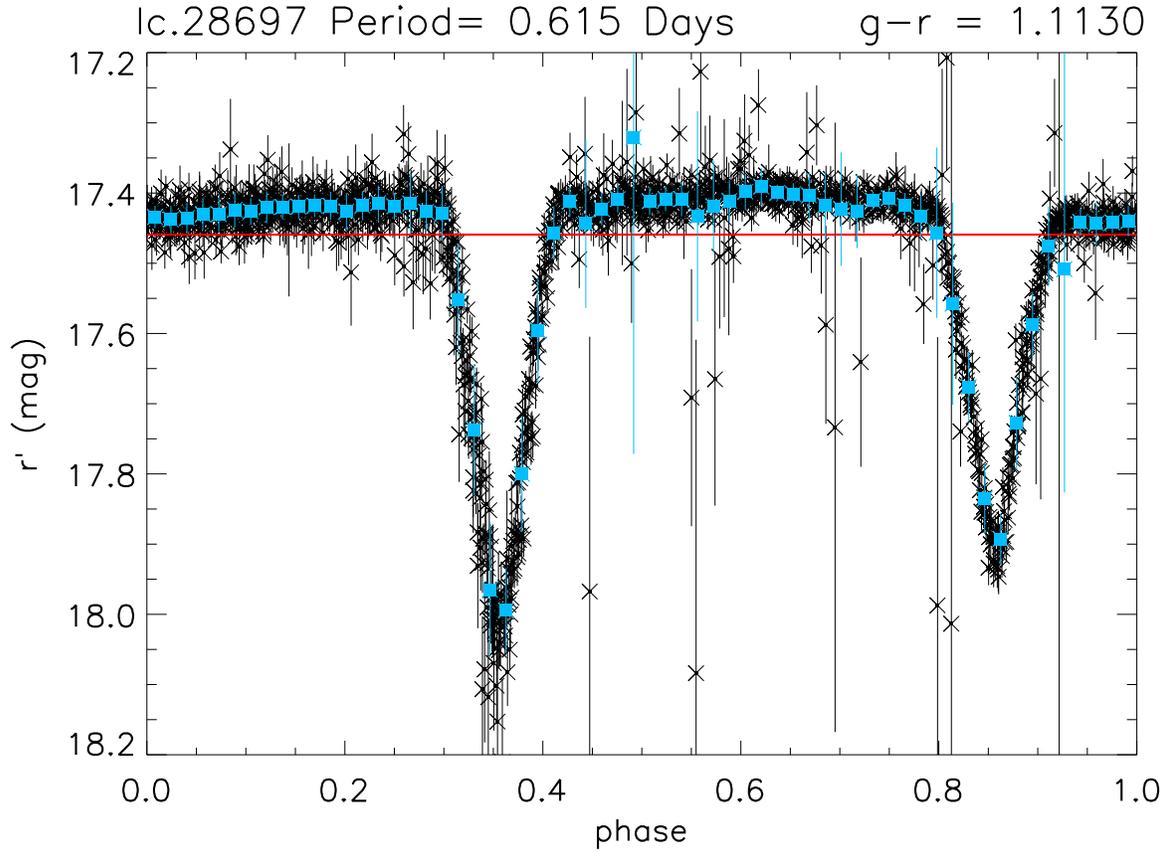}
\caption[ Eclipsing binary ]{\label{fig:eb2} This is an example of an
Algol-type eclipsing variable star that passed the initial OPTICSTAT tests for a transiting extrasolar
planet, but was removed after visual inspection.  The very deep (0.6 magnitude) and 
V-shaped primary eclipse and the clear presence of a secondary eclipse rules this
object out as a planetary transit system.  The symbols are the same as in Figure~\ref{fig:rrlyrae}. }
\end{figure}

\clearpage

\begin{figure}
\figurenum{20}
\plotone{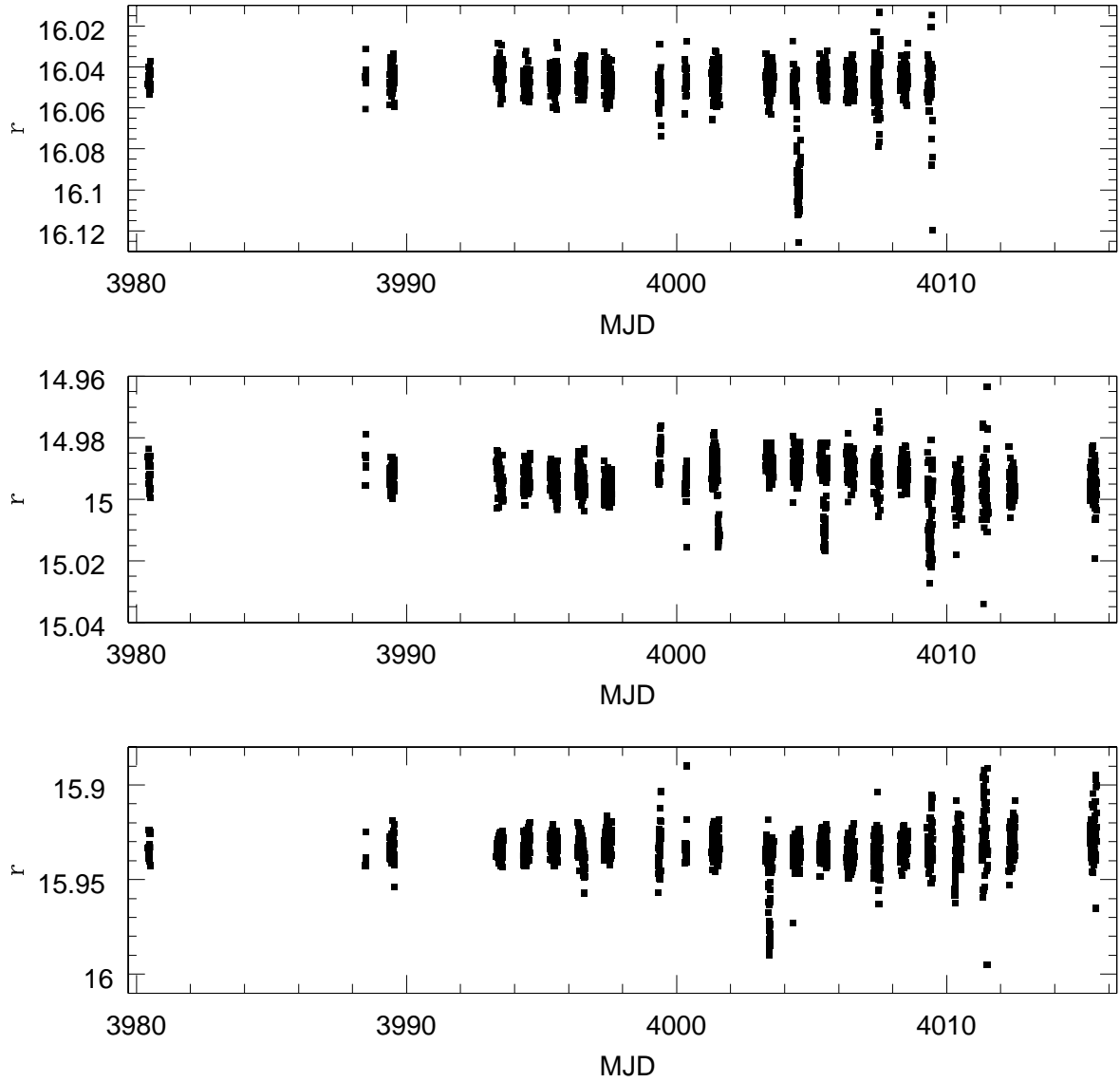}
\caption[ Final Candidates ]{\label{fig:candidates} The three highest
quality extrasolar candidates from the BOKS survey.  In order from 
top to bottom, they are BOKS-45069, BOKS-40959, and BOKS-52481.
The errorbars are omitted for clarity, but typical 2$\sigma$ uncertainties 
for these three light curves are 0.012, 0.014, 0.012 magnitudes respectively.}
\end{figure}

\begin{figure}
\figurenum{21}
\plotone{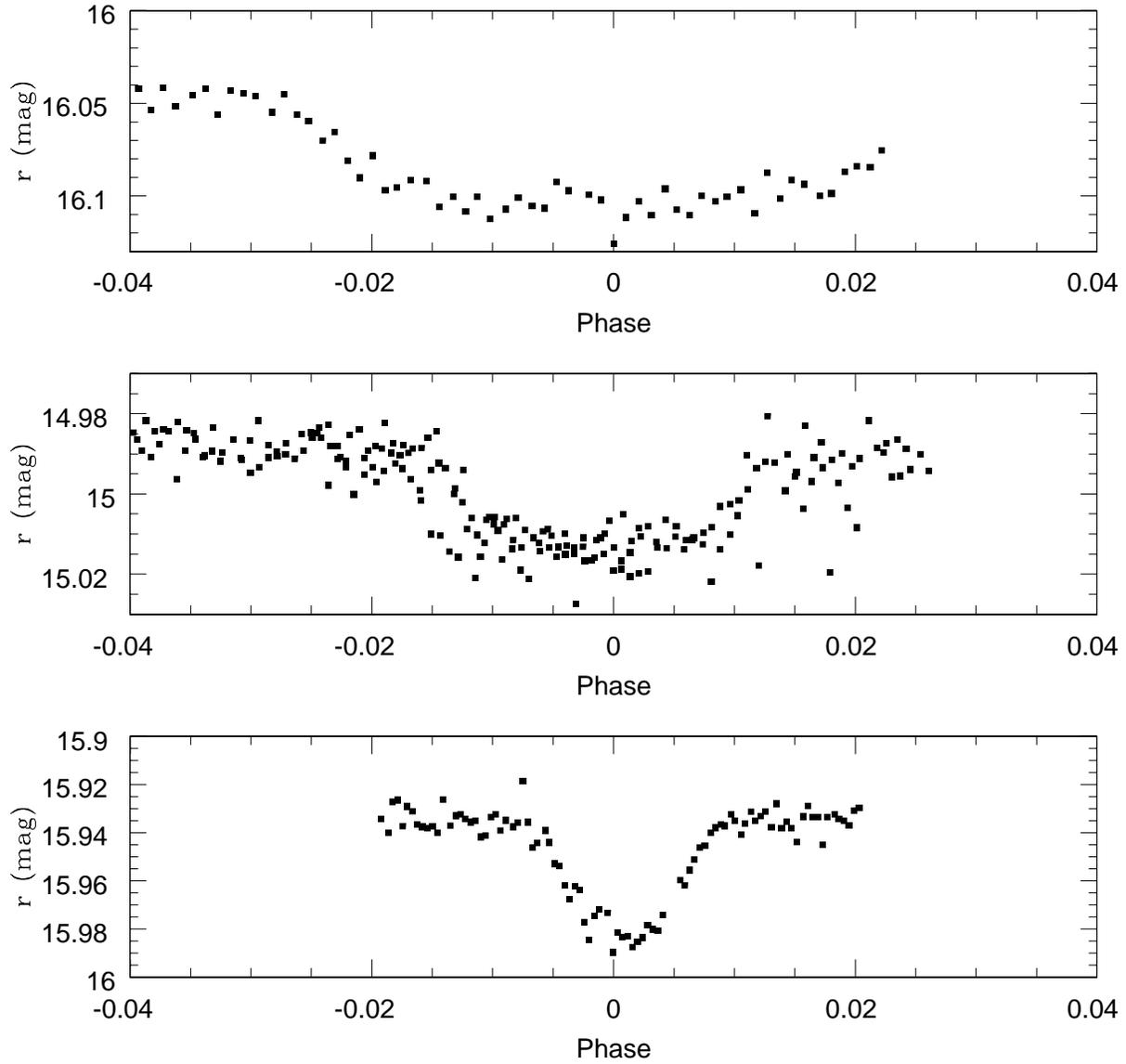}
\caption[ Final Candidates ]{\label{fig:candphase} The lightcurves
of our three highest quality extrasolar candidates near the transiting phase.
In order from top to bottom, they are BOKS-45069, BOKS-40959, and BOKS-52481.
The errorbars are omitted for clarity, but typical 2$\sigma$ uncertainties 
for these three light curves are 0.012, 0.014, 0.012 magnitudes respectively.}
\end{figure}

\begin{figure}
\figurenum{22}
\plotone{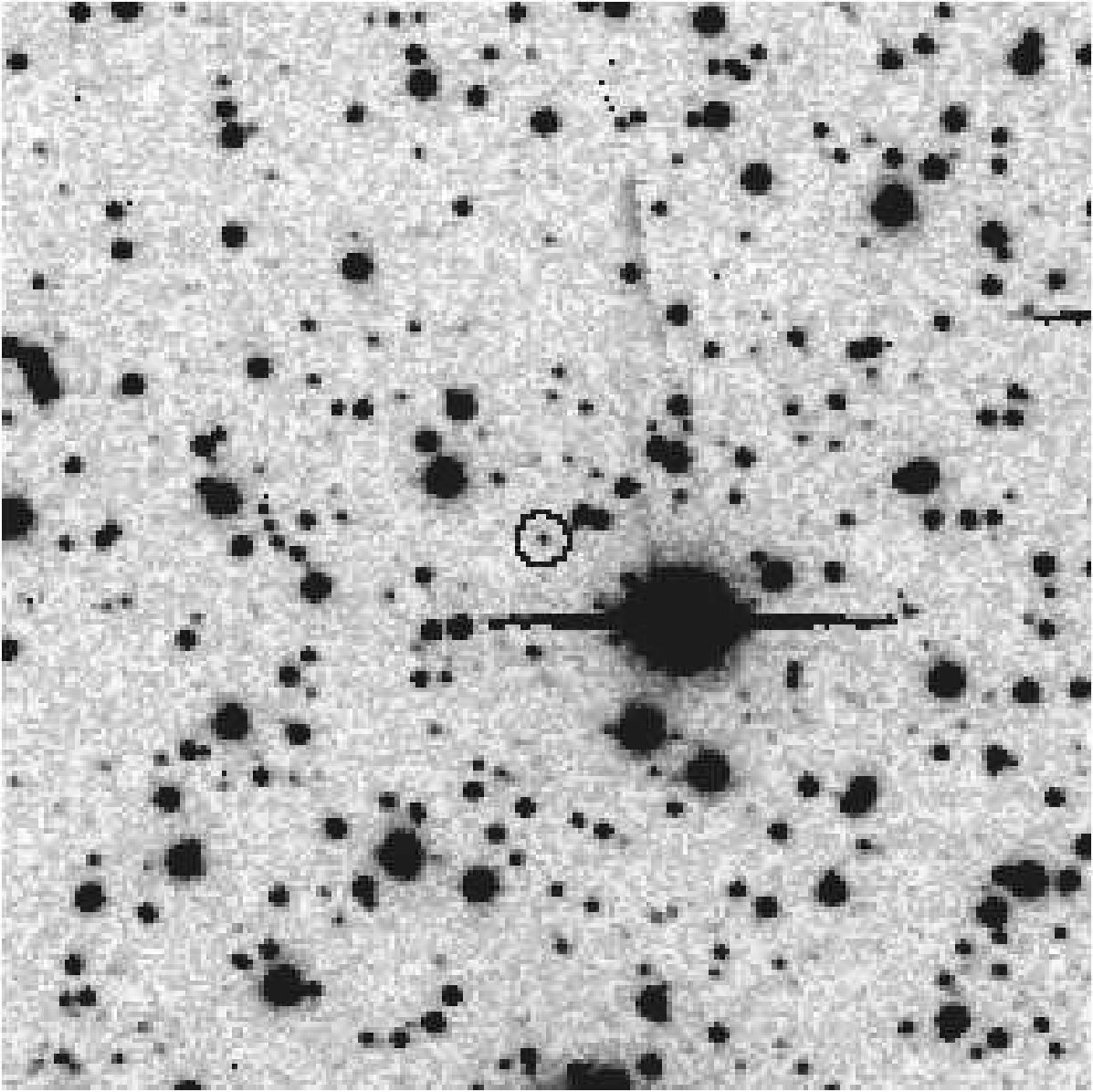}
\caption[ Dwarf Nova Finding Chart ]{\label{fig:dwarfnfc} The finding
chart of a dwarf nova found by the BOKS survey.
This finding chart is 5\arcmin $\times$ 5\arcmin ~in size, and North is up and
East is to the left on this image.  The object, known
as BOKS 45906, is located at 
$\alpha = 19^{h}40^{m}16.22^{s}$, $\delta = +46^{d}32^{m}48.23^{s}$,
and is centered in this image, surrounded by a circle for reference.
The emission north by northwest of this candidate is a Schmidt ghost, and
is unrelated to the source.  See \S\ref{sec:othervar} for a discussion of this object.}
\end{figure}

\begin{figure}
\figurenum{23}
\plotone{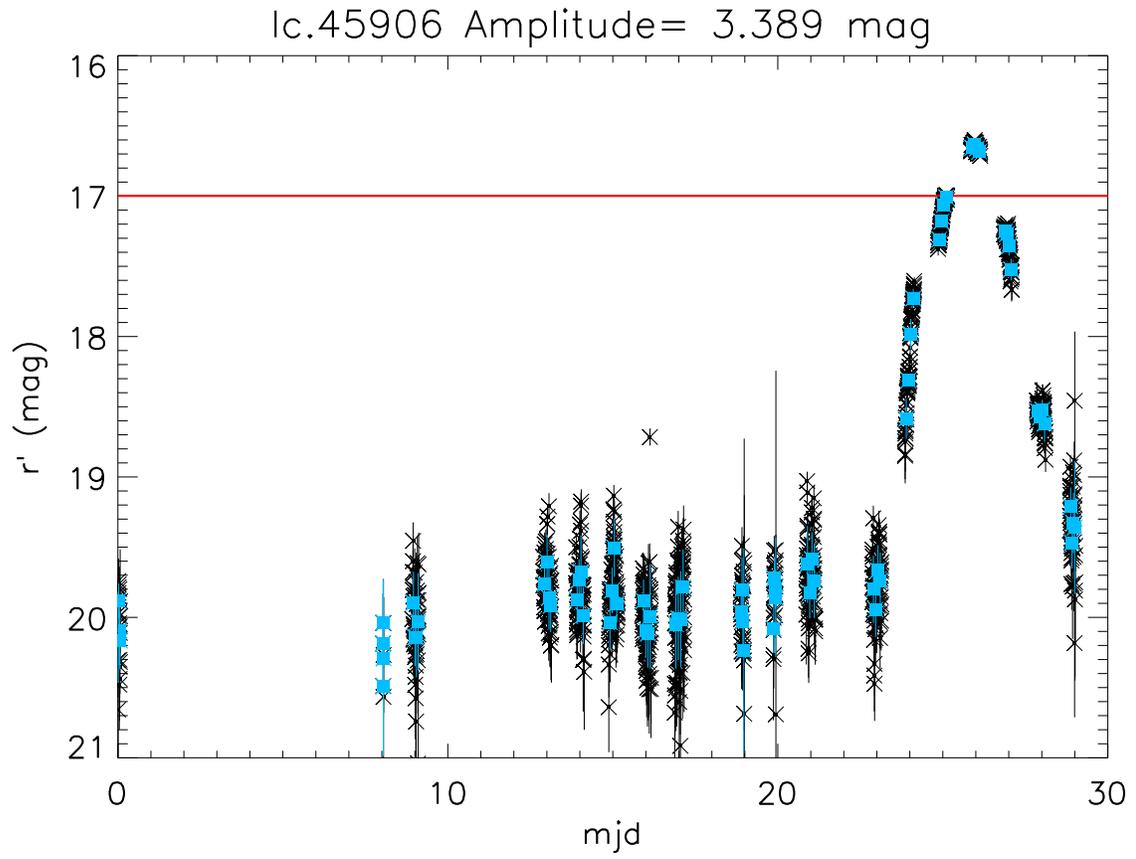}
\caption[ Dwarf Nova Light curve ]{\label{fig:dwarfn} The
light curve of a highly variable star found by the BOKS survey.  This
star, BOKS-45906, underwent a large photometric outburst, which
lasted approximately five days.  This behavior is typical of
dwarf novae (see \S\ref{sec:othervar} for discussion). }
\end{figure}

\clearpage

\begin{figure}
\figurenum{24}
\plotone{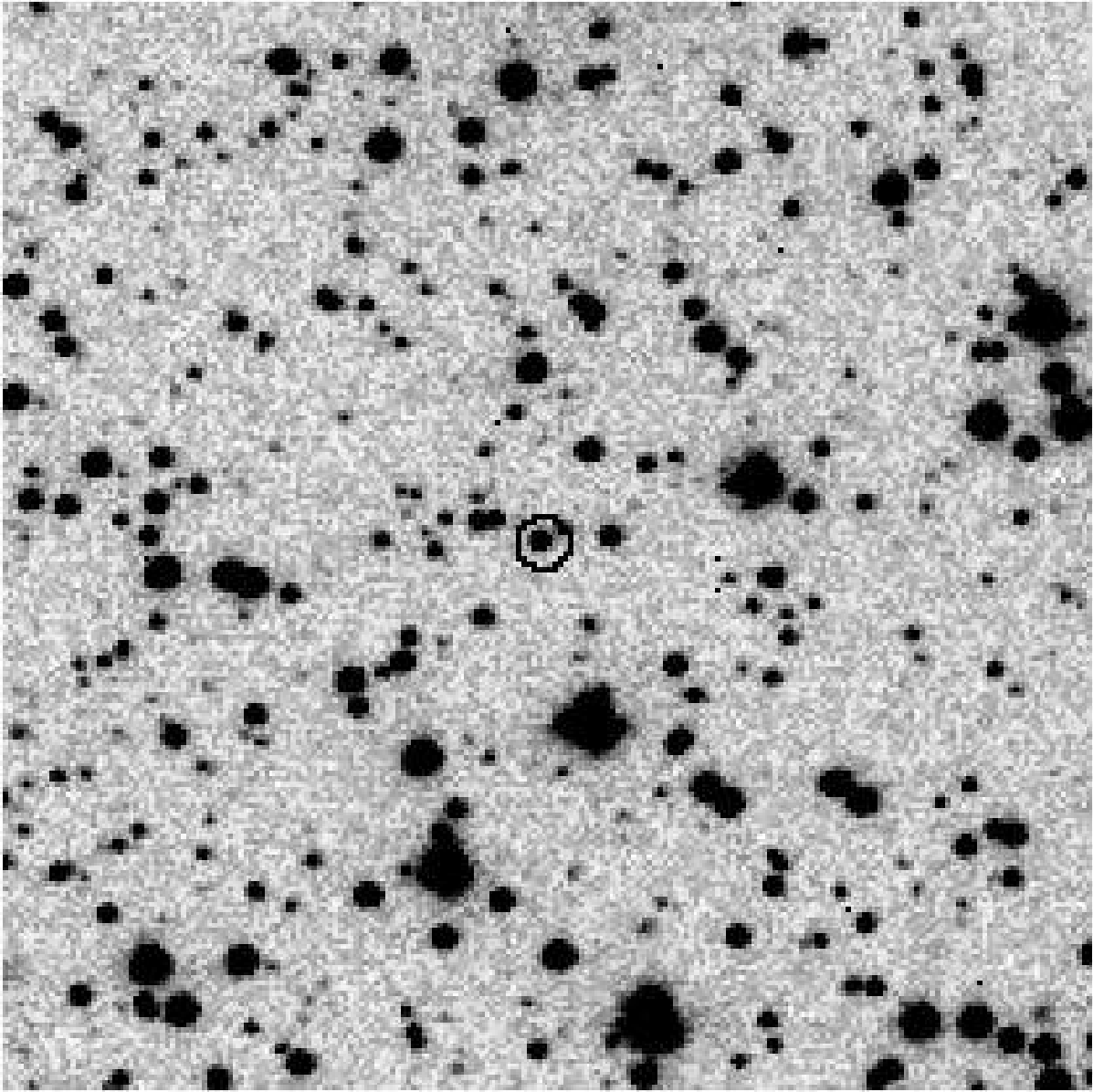}
\caption[ Blue Object Finding Chart ]{\label{fig:bluefc} The finding
chart for a very blue object found in the BOKS survey.  
This finding chart is 5\arcmin $\times$ 5\arcmin ~in size, and North is up and
East is to the left on this image.  The star, known as
BOKS-53856, is located at 
$\alpha = 19^{h}41^{m}31.35^{s}$, $\delta = +46^{d}06^{m}11.16^{s}$ 
and is centered in this image, surrounded by a circle for reference.
See \S\ref{sec:othervar} for a discussion of this object.}
\end{figure}

\begin{figure}
\figurenum{25}
\plotone{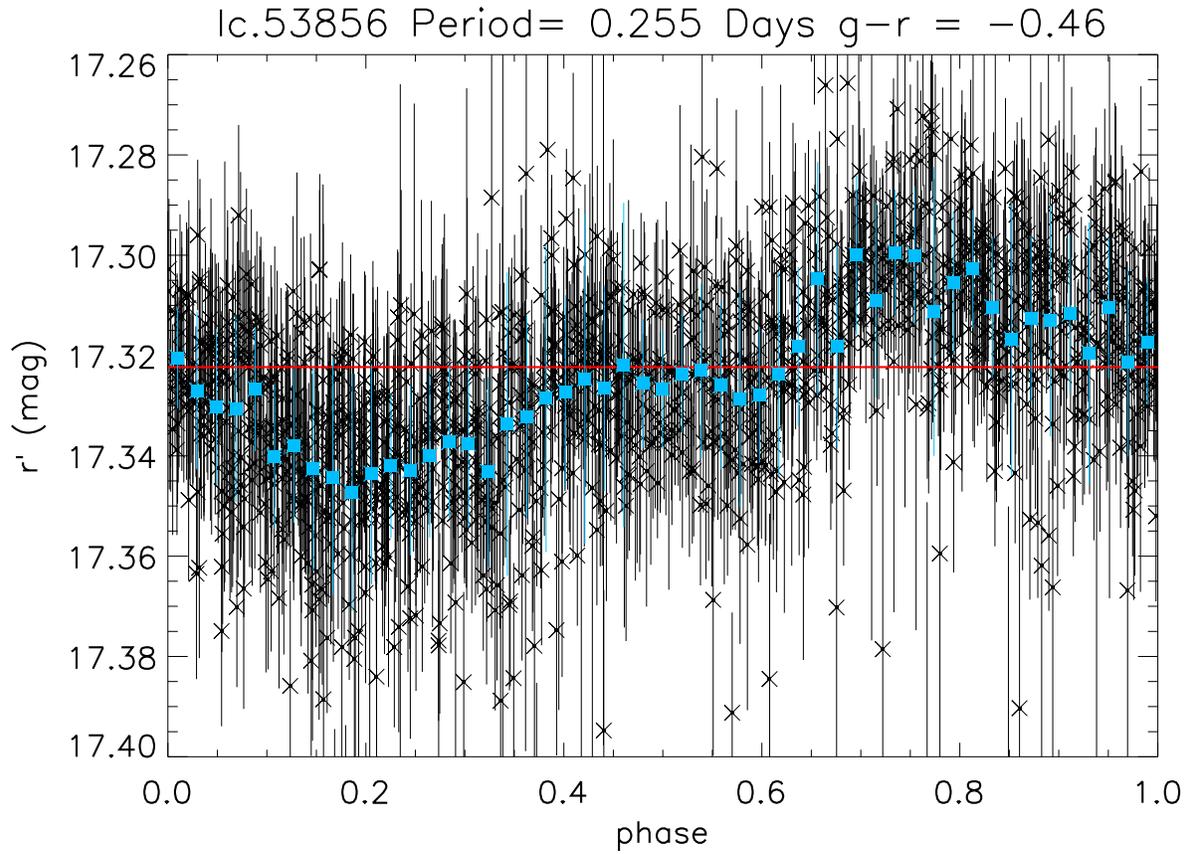}
\caption[ Blue object Light curve ]{\label{fig:bluelc} 
The phased light curve of an extremely blue ($g-r$ = -0.46) object in the 
BOKS field. The symbols are identical to Figure~\ref{fig:rrlyrae}.  
The light curve shows signs of periodic behavior, but with
unusual structure.  See \S\ref{sec:othervar} for further discussion. }
\end{figure}

\begin{figure}
\figurenum{26}
\plotone{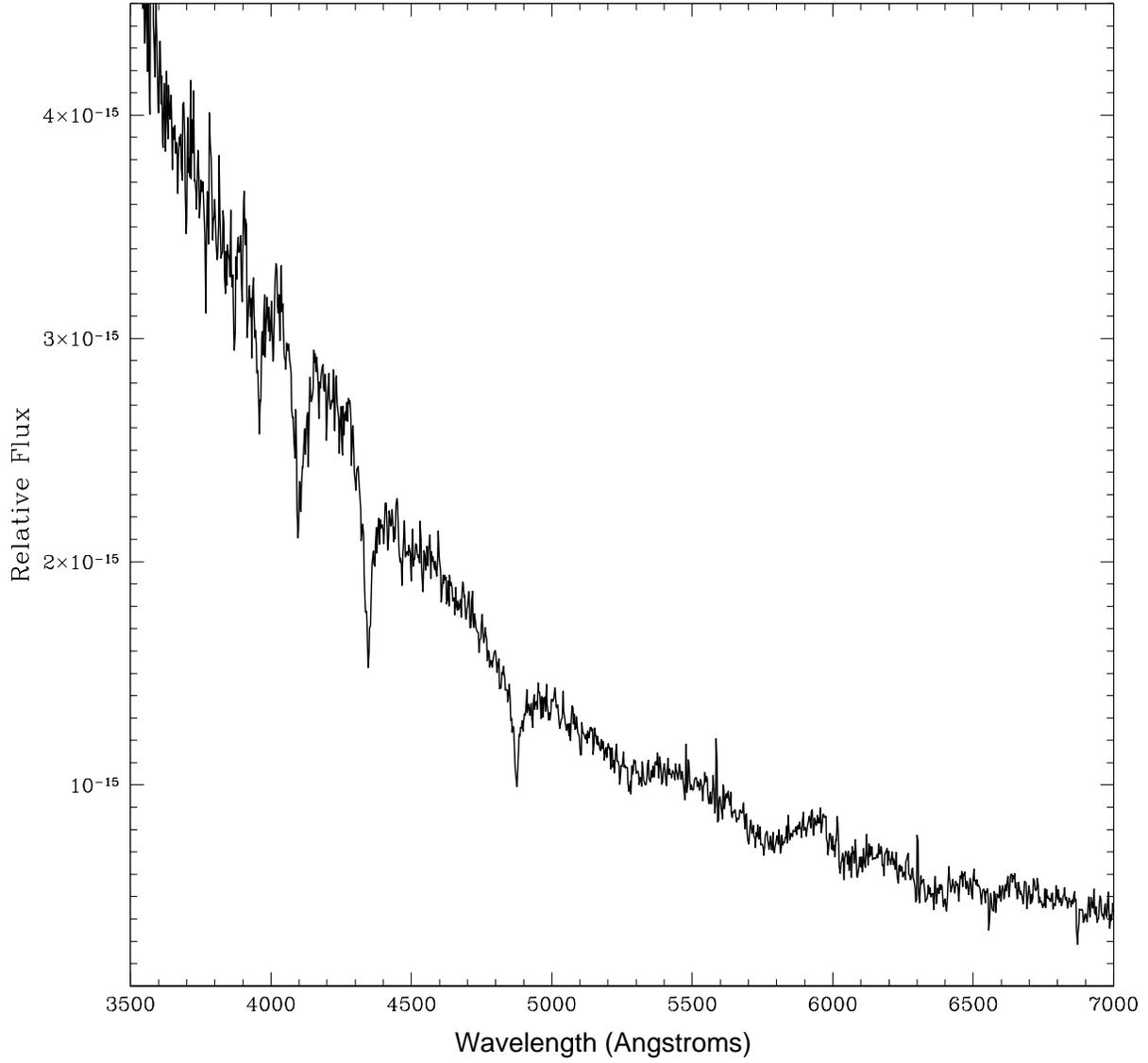}
\caption[ Blue object spectrum]{\label{fig:bluespec} 
A spectrum of BOKS-53856, a very blue object in the BOKS survey, obtained 
with the KPNO 2.1m telescope. Broad Balmer line features and a blue continuum 
are clearly visible, suggesting that this object is a white dwarf star.  
See \S\ref{sec:othervar} for further discussion.}
\end{figure}

\end{document}